\RequirePackage{rotating}
\PassOptionsToPackage{usenames,dvipsnames}{xcolour}
\documentclass[twocolumn,twocolappendix,iop]{openjournal}

\usepackage{hyperref}

\usepackage{amsmath,amstext}
\usepackage{ae,aecompl}
\usepackage[utf8]{inputenc}
\usepackage[figure,figure*]{hypcap}

\usepackage{url}
\usepackage{mdwlist}
\usepackage{listings}
\usepackage{rotating}
\usepackage{multirow}
\usepackage{newtxtext,newtxmath}
\usepackage{xspace}
\usepackage{natbib}
\usepackage[usenames,dvipsnames,svgnames,table]{xcolor}
\hypersetup{colorlinks,linkcolor=blue,citecolor=blue,urlcolor=violet}

\usepackage[a4paper,marginpar=10pt,headheight=20pt, right=1cm, top=1.5in, left=1in, bottom=0pt, footskip=0pt, footnotesep=10pt]{geometry}

\renewcommand{\arraystretch}{1.2}

\usepackage[T1]{fontenc}

\DeclareRobustCommand{\VAN}[3]{#2}
\let\VANthebibliography\thebibliography
\def\thebibliography{\DeclareRobustCommand{\VAN}[3]{##3}\VANthebibliography}

\DeclareRobustCommand{\PRF}[3]{#2} 
\newcommand{\inprep}{\emph{in preparation}}
\newcommand{\blockfont}[1]{{\textsc{#1}}\xspace}

\usepackage{graphicx}	
\usepackage{amsmath}	

\newcommand{\omegam}{\ensuremath{\Omega_\mathrm{m}}}
\newcommand{\omegab}{\ensuremath{\Omega_\mathrm{b}}}

\newcommand{\as}{\ensuremath{A_\mathrm{s}}}
\newcommand{\ns}{\ensuremath{n_\mathrm{s}}}
\newcommand{\lcdm}{$\Lambda$CDM}
\newcommand{\wcdm}{$w$CDM}

\begin{document}

\title{Joint constraints from cosmic shear, galaxy-galaxy lensing and galaxy clustering: \\ internal tension as an indicator of intrinsic alignment modelling error \vspace{-1.5cm}}

\author{S. Samuroff$^{1*}$}
\author{A. Campos$^{2,3}$}
\author{A. Porredon$^{4,5}$}
\author{and J. Blazek$^{1}$ \vspace{0.2cm}}

\affiliation{$^1$ Department of Physics, Northeastern University, Boston, MA, 02115, USA}
\affiliation{$^2$ McWilliams Center for Cosmology, Department of Physics, Carnegie Mellon University, Pittsburgh, PA 15213, USA}
\affiliation{$^3$ NSF AI Planning Institute for Physics of the Future, Carnegie Mellon University, Pittsburgh, PA 15213, USA}
\affiliation{$^{4}$ Ruhr University Bochum, Faculty of Physics and Astronomy, Astronomical Institute (AIRUB), German Centre for Cosmological Lensing, 44780 Bochum, Germany}
\affiliation{$^{5}$ Institute for Astronomy, University of Edinburgh, Royal Observatory, Blackford Hill, Edinburgh, EH9 3HJ, UK}

\email[$^*$]{s.samuroff@northeastern.edu}

\begin{abstract}
In cosmological analyses it is common to combine different types of measurement from the same survey. In this paper we use simulated Dark Energy Survey Year 3 (DES Y3) and Rubin Observatory Legacy Survey of Space and Time Year 1 (LSST Y1) data to explore the differences in sensitivity to intrinsic alignments (IA) between cosmic shear and galaxy-galaxy lensing data. We generate mock shear, galaxy-galaxy lensing and galaxy clustering data, contaminated with a range of IA scenarios. Using a simple 2-parameter IA model (NLA) in a DES Y3 like analysis, we show that the galaxy-galaxy lensing + galaxy clustering combination ($2\times2$pt) is significantly more robust to IA mismodelling than cosmic shear ($1\times2$pt). IA scenarios that produce up to $5\sigma$ biases in the $1\times2$pt case are seen to be unbiased at the level of $\sim1\sigma$ for $2\times2$pt. We demonstrate that this robustness can be largely attributed to the redshift separation in galaxy-galaxy lensing, which provides a cleaner separation of lensing and IA contributions. We identify a number of secondary factors which may also contribute, including the possibility of cancellation of higher-order IA terms in $2\times2$pt and differences in sensitivity to physical scales. Unfortunately this does not typically correspond to equally effective self-calibration in the $3\times2$pt analysis of the same data, which can show significant biases driven by the cosmic shear part of the data vector. If we increase the precision of our mock analyses to a level roughly equivalent to LSST Y1, we find a similar pattern with considerably more bias in a cosmic shear analysis than a $2\times2$pt one, and significant bias in the joint analysis of the two. Our findings suggest that IA model error can manifest itself as internal tension between $\xi_\pm$ and $\gamma_t + w$ data vectors. We thus propose that such tension (or the lack thereof) can be employed as a test of model sufficiency or insufficiency when choosing a fiducial IA model, alongside other data-driven methods.
\end{abstract}

\keywords{\centering cosmological parameters -- gravitational lensing: weak -- methods: statistical}

\maketitle

\section{Introduction}\label{sec:intro}

The study of weak lensing as a cosmological probe has evolved considerably in the last few years and decades. Although we talk about ``weak lensing" fairly loosely, the term actually encompasses several distinct measurements. Cosmic shear is the correlation of galaxy shapes with each other. That is, light from galaxies on adjacent lines of sight must pass through a similar cross-section of the Universe, and so the lensing distortions are correlated. It has been known for some time that shear-shear two-point correlations are a useful way to learn about cosmology \citep{bartelmann01,huterer02,hu04,frieman08}. Indeed, to date, cosmic shear analyses (referred to as $1\times2$pt analyses) have been key results of almost all galaxy imaging cosmology surveys (\citealt{benjamin07,kilbinger13, heymans13, jee16, hildebrandt17, troxel18, chang19, hikage19, hamana20, asgari21, loureiro22}; \citealt*{y3-cosmicshear2}; \citealt{y3-cosmicshear1,doux22,longley22,li23,dalal23,deskids23}).

Alternatively, instead of trying to detect lensing caused by background large scale structure, we can measure weak lensing around specific foreground lenses. One can do this with massive clusters, and so study their density profiles and total mass \citep{schneider00,king01}. Alternatively, one can use foreground galaxies as lenses, a measurement known as galaxy-galaxy lensing. Since the clustering pattern of galaxies traces out the large scale structure of the Universe, these shear-position correlations measure a similar physical effect to shear-shear ones. If we have a good estimate for the galaxy-to-dark matter mapping (i.e. galaxy bias) and the distribution of redshifts (or better yet, precise redshifts for individual galaxies), we can use galaxy-galaxy lensing to probe the properties of the Universe. There is a relatively long history of this, using both spectroscopic and photometric surveys (\citealt{sheldon04,baldauf10,mandelbaum13,kwan17}; \citealt*{y1-ggl}; \citealt{alam17,leauthaud17,yoon19,blake20,singh20,miyatake21,lee22,y3-ggl,y3-2x2pt_maglim,y3-2x2pt_redmagic}). Often studies of this sort (including many of those cited above) will also incorporate galaxy clustering auto-correlations (\citealt{cole05,blake12,aubourg15,y1-clustering,alam21,zhou21,y3-clustering}; \citealt*{sanchez22}), in order to break the degeneracy between galaxy bias and the clustering amplitude $\sigma_8$. This combination is often referred to as $2\times2$pt.

It is also worth noting briefly that there is a subfield of weak lensing studies looking at the cross-correlation between galaxy surveys (lensing and positions) and Cosmic Microwave Background (CMB) lensing. These are complementary to other types of weak lensing (\citealt{namikawa19,marques20,robertson21,krolewski21,omori22,chang22,y3-6x2pt}). The nomenclature of the combinations beyond $3\times2$pt is slightly less well-defined, but CMB lensing is not the focus of this work, so we will not go into the details here. 

Although cosmic shear and galaxy-galaxy lensing are powerful probes in their own right, combining the two with galaxy clustering in a joint analysis ($3\times2$pt) has become an increasingly mainstream part of modern cosmology. This is for a number of reasons, not least that the two have different sensitivities to cosmological parameters, and so together can break internal degeneracies \citep{hu04,frieman08,cacciato09,yoo12}. Another factor is the ability of joint data vectors to self-calibrate photometric redshift error and other systematic uncertainties \citep{huterer06,bridle07,bernstein09,joachimi09,samuroff17}. In the past five or so years, joint-probe analyses using cosmic shear, galaxy-galaxy lensing and galaxy clustering together have provided some of the most powerful late-universe cosmology constraints to date \citep{y1-kp,vanUitert18,joudaki18,heymans21,y3-kp,miyatake23,sugiyama23}.

Before combining any given set of measurements, however, it is usual to demonstrate their consistency. If two data sets are found to be discrepant, it points either to systematics in one or the other, or the need for a new model. This can be an interesting new result or a nuisance, but either way, one should not combine the discrepant data sets. Judging consistency using projected confidence contours can be misleading, but fortunately a range of metrics using the full parameter space have been proposed. For uncorrelated data (e.g. from non-overlapping surveys or very different observables), various statistics have been developed, both Bayesian evidence- and likelihood-based (\citealt*{y3-tensions}; \citealt{raveri19}). For probes that are correlated, such as cosmic shear and galaxy-galaxy lensing measurements from the same survey, alternatives are commonly used. For example, the DES Y3 analysis used a method based on the Posterior Predictive Distribution (PPD), as described in \citet*{y3-internal_tensions}.

It has long been known that an effect known as intrinsic alignment (IA) has the potential to bias weak lensing based measurements. IA is the name given to correlations arising from the fact that the \emph{intrinsic} (i.e. pre gravitational shear) shapes of galaxies are not entirely random. This induces correlations both between physically close pairs (known as intrinsic-intrinsic or II correlations) and between the shear of one galaxy and the intrinsic shape of another (known as shear-intrinsic or GI correlations). A related but slightly different correlation also arises between the clustering density and intrinsic shapes of close-by objects ($g$I). 

The reason IAs can cause bias is simple: they introduce new features into two-point lensing data, which appear similar to lensing (although not identical, allowing self-calibration of the sort discussed below). If the IA model used to analyse that data cannot match them perfectly, then other parameters may need to adjust in order to maintain a good overall fit. Sometimes this can happen inside the IA model space, in which case the mapping between IA parameters and underlying physical processes becomes more complicated. However, unmodelled IAs can quite can easily (if imperfectly; see \citealt{campos22}) be absorbed by cosmological parameters too (\citealt{krause16,blazek19,fortuna21}; \citealt*{y3-cosmicshear2}; \citealt{campos22}). This subject, and the question of IA model sufficiency, has been discussed relatively widely in the context of cosmic shear. One aspect that has received somewhat less attention is how parameter biases play out in the context of a multi-probe analysis. That is, how different the sensitivities of galaxy-galaxy lensing and cosmic shear are to IA model error, and how effectively the combination can  mitigate such error. In this work we consider exactly this question, using simulated Dark Energy Survey Year 3 (DES Y3) and Legacy Survey of Space and Time Year 1 (LSST Y1) like data. 

This paper is structured as follows. In Section \ref{sec:theory} we discuss how our joint data vector is modelled. We pay specific attention to the modelling of the various IA terms, since these are important for our conclusions. Section \ref{sec:data} then considers a number of simulated setups with different lens and source sample configurations. We also set out the details of the mock analyses, including the nuisance parameters associated with each sample. In Section \ref{sec:results} we discuss our main results and test their robustness to variations in the analysis and mock data. We conclude in Section \ref{sec:conclusions}. 

\vspace{0.25cm}
\section{Modelling}\label{sec:theory}

This section describes the modelling framework used in this paper. In short, we use a DES Y3 like $3\times2$pt analysis, combining cosmic shear, galaxy-galaxy lensing and galaxy clustering. Although we highlight the key features relevant for this work, much of the infrastructure was developed for the DES Y3 analyses and is documented more extensively in \citet{y3-methods} and the accompanying DES papers. For part of our results, we also consider an LSST Y1 like analysis. We use many of the same tools for this, but with key modifications that we will highlight. We consider below the details of how $\xi_\pm$ and $\gamma_t$ are calculated, since these are both sensitive to IAs, but in slightly different ways -- a point that will be important later. For details about how $w(\theta)$ is computed, see the papers cited above.

\subsection{Cosmic shear}\label{sec:theory:shear_gg}

Cosmic shear is the name given to small distortions in the observed shapes of galaxies due to weak gravitational lensing by foreground large scale structure. At a particular point on the sky, the shear $\gamma^{\rm G}$ can be written as a projection of the matter overdensity field $\delta$ (or, rather, the spatial derivatives of $\delta$) along the line-of-sight. Since light from galaxies that appear close together on the sky trace similar paths through the Universe, $\gamma^{\rm G}$ is spatially correlated. Indeed, two-point functions, which measure the shear-shear correlation $\left \langle \gamma^{\rm G} \gamma^{\rm G} \right \rangle$ as a function of angular separation $\theta$, are the most common way to observe cosmic shear. Since shear is a spin-2 quantity, there are two different cosmic shear correlations that can be constructed, $\xi_+$ and $\xi_-$. These are sensitive to different physical scales, and so are complimentary to each other (see e.g. \citealt*{y3-cosmicshear2} Figure 2). Most commonly, these correlations are constructed between galaxies in redshift bins $i,j$, which is useful in order to break degeneracies and constrain the evolution of $\delta$ with redshift.

Starting with the 3D nonlinear matter power spectrum $P_\delta (k,z)$, and assuming the Limber approximation \citep{limber53,loVerde08}, we can write down an expression for the 2D angular power spectrum of convergence $\kappa$ as: \vspace{-0.5cm}

\begin{multline}\label{eq:C_GG}
    C^{ij}_\kappa(\ell) = \int_0^{\chi_{\rm hor}} \mathrm{d}\chi 
 \frac{W_{\rm s}^i(\chi) W_{\rm s}^j(\chi)}{\chi^2}  \\ \times P_\delta \left ( k=\frac{\ell+0.5}{\chi}, z(\chi) \right ).
\end{multline}

\noindent
We have defined a few things here. First $\chi$ is a comoving distance from the observer and $z(\chi)$ is the corresponding redshift (assuming some cosmology). The integral limit $\chi_{\rm hor}$ is the comoving horizon distance. The lensing kernel $W_{\rm s}^i$ is given by:

\begin{equation}
    W_{\rm s}^{i}(\chi) = \frac{3}{2c^2} H^2_0 \omegam \frac{\chi}{a(\chi)} \int^{\chi_{\rm hor}}_{\chi} \mathrm{d}\chi' n_{\rm s}^i\left (z(\chi) \right ) \frac{\chi' - \chi}{\chi'} \frac{\mathrm{d}z}{\mathrm{d}\chi'}.
\end{equation}

\noindent
The source redshift distribution $n_{\rm s}(z)$ is normalised to integrate to 1. For an illustration of what $n_{\rm s}$ and $W_{\rm s}$ look like as a function of redshift for the fiducial analyses presented in this paper, see the upper two panels of Figure \ref{fig:nofz}.

The conversion from angular multipoles $\ell$ to separation $\theta$ can be written as:

\begin{align}
\nonumber    \xi^{ij}_+ (\theta) = \sum_\ell \frac{2\ell+1}{2\pi \ell^2 (\ell+1)^2}  \left ( G^{+}_{\ell,2}(\mathrm{cos} \theta) + G^{-}_{\ell,2}(\mathrm{cos} \theta) \right )  \\
    \times \left ( C^{ij}_{\gamma, \mathrm{EE}} (\ell) + C^{ij}_{\gamma, \mathrm{BB}} (\ell) \right ),
\end{align}

\begin{align}
\nonumber    \xi^{ij}_- (\theta) = \sum_\ell \frac{2\ell+1}{2\pi \ell^2 (\ell+1)^2}  \left ( G^{+}_{\ell,2}(\mathrm{cos} \theta) - G^{-}_{\ell,2}(\mathrm{cos} \theta) \right ) \\
     \times \left ( C^{ij}_{\gamma, \mathrm{EE}} (\ell) - C^{ij}_{\gamma, \mathrm{BB}} (\ell) \right )
\end{align}

\noindent
(see \citealt{stebbins96}).
In the above $G^{+}_{\ell,2}$ and $G^{-}_{\ell,2}$ are bin-averaged functions defined in \citet{y3-methods}. In the simplest case, the convergence and E-mode shear spectra are equivalent here, $C_{\gamma,\mathrm{EE}} \rightarrow C_\kappa$ and $C_{\gamma,\mathrm{BB}}=0$. In practice, however, intrinsic alignments can contribute to both the E- and B-mode $C_\gamma$ spectra (see Section \ref{sec:theory:ias}).

For a given set of cosmological parameters, we evaluate $P_\delta(k,z)$ using the Boltzmann code \blockfont{CAMB} (for the linear part; \citealt{lewis00}) and \blockfont{HaloFit} (for the nonlinear corrections; \citealt{takahashi12}). Note that unlike \citet*{y3-cosmicshear2} and \citet{y3-cosmicshear1}, our fiducial results do not make use of shear ratios (ratios of galaxy-galaxy lensing measurements on small scales, which were included as an additional likelihood in the fiducial Y3 analyses; see \citealt*{y3-sr}).

\subsection{Galaxy-galaxy lensing}\label{sec:theory:ggl}

Galaxy-galaxy lensing is a similar measurement to cosmic shear, described above. Instead of the auto-correlation of shear, however, we are now measuring the cross-correlation with galaxy density $\left \langle \delta_{g} \gamma^{\rm G} \right \rangle$. One can write down the equivalent transformation to Eq. \eqref{eq:C_GG}, again assuming the Limber approximation, as:\vspace{-0.25cm}

\begin{multline}
    C^{ij}_{\delta_g \kappa}(\ell) = \int_0^{\chi_{\rm hor}} \mathrm{d}\chi 
 \frac{n_\mathrm{l}^i(\chi) W_{\rm s}^j(\chi)}{\chi^2} \\ \times P_{\delta_g \delta} \left ( k=\frac{\ell+0.5}{\chi}, z(\chi) \right ).
\end{multline}

\noindent
As before, $n_{\rm l}(z)$ is a redshift distribution that integrates to 1, but this time of the lens galaxies. We model the galaxy-matter 3D power spectrum $P_{\delta_g \delta}$ by assuming linear bias, such that:

\begin{equation}
    P_{\delta_g \delta}(k,z) = b_1 P_{\delta}(k,z).
\end{equation}
\noindent
The linear bias coefficient $b_1$ is expected to depend on the details of the sample and to have some evolution with redshift. In our fiducial setup, we thus allow one independent parameter $b_1^i$ for each lens redshift bin $i$, which are marginalised with uncorrelated uninformative priors (see Section \ref{sec:data:3x2pt} for discussion). 

To test our results in Section \ref{sec:results}, a subset of our analyses use a nonlinear bias model, which is based on a perturbative expansion to third order (see \citealt{pandey22}). In short, this includes three extra parameters: $b_2$, $b_{s^2}$ and $b_{\rm 3nl}$. For the sake of simplicity (in order to allow the nonlinear bias model to reproduce mock data containing linear bias), we fix the latter two parameters to zero, rather than their non-zero coevolution values (see \citealt{pandey22} Sec. 2D). We also follow \citet{y3-2x2pt_maglim} and \citet{y3-2x2pt_redmagic} in choosing to vary the product $\sigma_8  b_1$ and $\sigma_8^2  b_2$ rather than $b_1$ and $b_2$ directly. This is a practical decision to limit projection effects\footnote{Projection effects are shifts in marginalised posteriors that can arise from the projection of a multidimensional volume down to one or two axes. These are not biases in the usual sense, since they do not indicate any sort of model/data mismatch, and there are statistics (e.g. the global \emph{Maximum a Posteriori}) that are unaffected. See e.g. \citealt{y3-methods} Sec IV-A for discussion.}, but is not expected to otherwise affect the results. This amounts to one extra free parameter per lens bin. We vary the bias parameters with wide flat priors $\sigma_8 b^i_1 = [0.67,3.0]$ and $\sigma_8^2 b^i_2 = [-4.2, 4.2]$. Note that use of this model is limited to the tests in Section \ref{sec:results:impact_of_bias_model} -- for all other chains we use the linear bias model.

It is also worth mentioning that, in principle, cross-terms arise between nonlinear galaxy bias and higher-order IA contributions. These are expected to be small, and so they were neglected in the modelling pipeline for DES Y3. Also note that no part of our analysis includes \emph{both} TATT and nonlinear galaxy bias; chains (and data vectors) run with the former assume linear bias and those run with the latter assume NLA.

With the angular power spectrum in hand, we can evaluate the real space galaxy-galaxy lensing correlation as:

\begin{equation}\label{eq:gammat}
    \gamma^{ij}_t(\theta) = \sum_\ell \frac{2\ell+1}{4\pi \ell(\ell + 1)} \bar{P}^{2}_\ell (\mathrm{cos} \theta) C^{ij}_{g \gamma}(\ell),
\end{equation}

\noindent
where $\bar{P}^2_\ell$ are associated Legendre polynomials, averaged within each $\theta$ bin (see e.g. \citealt{y3-ggl} Eq. 25). Similarly to cosmic shear above, in the absence of intrinsic alignments and magnification, $C_{g \gamma} \rightarrow C_{\delta_g \kappa}$ in Eq. \eqref{eq:gammat}.

\subsection{Intrinsic Alignments \& Magnification}\label{sec:theory:ias}

In practice, both cosmic shear and galaxy-galaxy lensing have contributions from effects other than pure lensing and galaxy clustering. First, IAs add a spatially correlated shape component, such that the observed shear is $\gamma = \gamma^{\rm G} + \gamma^{\rm I}$, giving rise to $\left \langle \gamma^{\rm G} \gamma^{\rm I} \right \rangle$ and $\left \langle  \gamma^{\rm I} \gamma^{\rm I} \right \rangle$ correlations at the two-point level (often referred to simply as GI and II). Analogously, the observed density of galaxy counts on a patch of sky is altered by magnification, $\delta_{g, \mathrm{obs}} = \delta_g + \delta_\mu$. When writing out the angular correlation functions, we have several additional correlations.

\begin{equation}
    C^{ij}_{\gamma, \mathrm{EE}} = C^{ij}_\kappa + C^{ij}_{\rm GI} + C^{ij}_{\rm IG} + C^{ij}_{\rm II, EE}
\end{equation}

\begin{equation}
    C^{ij}_{\gamma, \mathrm{BB}} = C^{ij}_{\rm II, BB}
\end{equation}

\begin{equation}\label{eq:C_g_gamma}
    C^{ij}_{g \gamma} = C^{ij}_{ \delta_g \kappa} + C^{ij}_{\delta_g \mathrm{I}} + C^{ij}_{\mu \kappa} + C^{ij}_{\mu \mathrm{I}}
\end{equation}

\noindent
Since, to first order, there is no B-mode contribution from lensing, $C^{ij}_{\gamma, \mathrm{BB}}$ is non-zero only due to intrinsic alignments\footnote{We are ignoring other theoretical sources of B-modes such as source clustering \citep{schneider02,schmidt09}, which are commonly assumed to be small (at least for two-point statistics -- see \citealt{y3-methods} and \citealt{gatti24} for discussion).}. The Limber integrals for each of the terms above can be expressed in a similar way to those in Sections \ref{sec:theory:shear_gg} and \ref{sec:theory:ggl}: 

\begin{multline}
    C^{ij}_{\rm GI}(\ell) = \int_0^{\chi_{\rm hor}} \mathrm{d}\chi 
 \frac{ W_{\rm s}^i(\chi) n_\mathrm{s}^j(\chi) }{\chi^2}  \\ \times P_{\rm GI} \left ( k=\frac{\ell+0.5}{\chi}, z(\chi) \right ),
\end{multline}

\begin{multline}
    C^{ij}_{\rm II, EE/BB}(\ell) = \int_0^{\chi_{\rm hor}} \mathrm{d}\chi 
 \frac{ n_{\rm s}^i(\chi) n_\mathrm{s}^j(\chi) }{\chi^2}  \\ \times P_{\rm II, EE/BB} \left ( k=\frac{\ell+0.5}{\chi}, z(\chi) \right ),
\end{multline}

\begin{multline}\label{eq:C_gI}
    C^{ij}_{\delta_g \mathrm{I}}(\ell) = \int_0^{\chi_{\rm hor}} \mathrm{d}\chi 
 \frac{ n_{\rm l}^i(\chi) n_\mathrm{s}^j(\chi) }{\chi^2}  \\ \times P_{\delta_g \mathrm{I}} \left ( k=\frac{\ell+0.5}{\chi}, z(\chi) \right ),
\end{multline}

\begin{multline}\label{eq:C_mI}
    C^{ij}_{\mu \mathrm{I}}(\ell) = \int_0^{\chi_{\rm hor}} \mathrm{d}\chi 
 \frac{ W_{\rm l}^i(\chi) n_\mathrm{s}^j(\chi) }{\chi^2}  \\ \times P_{\mu \mathrm{I}} \left ( k=\frac{\ell+0.5}{\chi}, z(\chi) \right ),
\end{multline}

\begin{multline}\label{eq:C_mG}
    C^{ij}_{\mu \kappa(\ell)} = \int_0^{\chi_{\rm hor}} \mathrm{d}\chi 
 \frac{ W_{\rm l}^i(\chi) W_\mathrm{s}^j(\chi) }{\chi^2}  \\ \times P_{\mu \delta} \left ( k=\frac{\ell+0.5}{\chi}, z(\chi) \right ).
\end{multline}

\noindent
The power spectra in the latter three can be related to the GI IA spectrum and the nonlinear matter power spectrum by assuming linear bias:\vspace{-0.25cm}

\begin{equation}\label{eq:P_gI}
    P_{\delta_g \mathrm{I}}(k,z) = b_1 P_{\rm GI}(k,z),
\end{equation}

\begin{equation}
    P_{\mu \mathrm{I}}(k,z) = C P_{\rm GI}(k,z),
\end{equation}

\begin{equation}
    P_{\mu \delta}(k,z) = C P_{\delta}(k,z),
\end{equation}

\noindent
with the factor $C$ being the magnification coefficient (see \citealt*{y3-magnification} for a definition and further discussion), which describes the overall impact of magnification\footnote{We follow \citealt*{y3-magnification} here in parameterising the overall impact of magnification with a single amplitude per redshift bin. In terms of physics, one has two competing effects due to the fact that magnification both boosts the observed fluxes of galaxies and also expands the apparent area of a given patch on the sky. In the notation of \citet{joachimi09}, $C_{\rm mag}=2(\alpha-1)$, where $\alpha$ is the logarithmic slope of the faint end of the luminosity function.}. Note that $C$ is not included as a free parameter in our analyses, but rather fixed to the fiducial values measured by \citealt*{y3-magnification} in each bin (see also Table \ref{tab:params}). As mentioned in the previous section, a subset of our chains include nonlinear galaxy bias. Note that the Y3 nonlinear bias model does not include an expansion of Eq. \ref{eq:P_gI} to include all relevant terms -- the model always assumes a linear relation between $P_{\delta_g \mathrm{I}}$ and $P_{\rm GI}$ (as in \citealt{y3-methods}).

Although we include it in all our modelling, the magnification-intrinsic term $\mu$I is expected to be much smaller in magnitude than $\delta_g$I, and the impact is thus expected to be minimal (see the turquoise dashed line in \citealt{y3-ggl}'s Figure 7). By considering the equations above, we can see that (with certain assumptions), all of the 2D angular spectra entering our cosmic shear and galaxy-galaxy lensing measurements are derived from four 3D power spectra: the matter power spectrum $P_\delta$, and three intrinsic alignment spectra $P_{\rm GI}$, $P_{\rm II, EE}$ and $P_{\rm II, BB}$.  

We have discussed how we estimate $P_\delta$ in Section \ref{sec:theory:shear_gg}. Using the formalism of \citet{blazek19}, one can write the three IA power spectra as:

\begin{equation}\label{eq:tatt_gi}
    P_{\rm GI} = C_1 P_\delta + b_{\rm TA} C_1 P_{0|0E} + C_2 P_{0|E2},
\end{equation}

\begin{multline}\label{eq:tatt_ii}
    P_{\rm II, EE} = C^2_1 P_\delta + 2 b_{\rm TA} C^2_1  P_{0|0E} + b^2_{\rm TA} C^2_1 P_{0E|0E} \\ + C^2_2 P_{E2|E2} + 
    2C_1C_2 P_{0|E2} + 2b_{\rm TA}C_1 C_2 P_{0E|E2},
\end{multline}

\begin{multline}\label{eq:tatt_ii_bb}
    P_{\rm II, BB} = b^2_{\rm TA} C^2_1P_{0B|0B} + C^2_2P_{B2|B2} \\ + 2b_{\rm TA} C_1 C_2 P_{0B|B2}.
\end{multline}

\noindent
We should note here that $C_{1,2}$ in Eq. \ref{eq:tatt_gi}-\ref{eq:tatt_ii_bb} are IA amplitudes, and are \emph{not} related to the magnification coefficients discussed earlier (despite the similar notation). The various scale dependent terms, $P_X$, can all be calculated to one-loop order as integrals of the linear matter power spectrum over $k$ \citep{blazek19}. We perform these integrals within \blockfont{CosmoSIS} using \blockfont{FastPT}\footnote{\url{https://cosmosis.readthedocs.io/en/latest/reference/standard_library/fast_pt.html}} \citep{mcewen16,fang17}. The amplitudes $C_1$ and $C_2$ are given by:

\begin{equation}\label{eq:tatt_a1}
    C_1(z) = -A_1  \frac{\bar{C}_1 \rho_{\rm c} \omegam}{D(z)} \left ( \frac{1+z}{1+z_0} \right )^{\eta_1},
\end{equation}

\begin{equation}\label{eq:tatt_a2}
    C_2(z) = 5 A_2  \frac{\bar{C}_1 \rho_{\rm c} \omegam}{D^2(z)} \left ( \frac{1+z}{1+z_0} \right )^{\eta_2}.
\end{equation}

\noindent
The pivot redshift $z_0$ is set to $z_0=0.62$ and the constant $\bar{C}_1$ is fixed at a value of $\bar{C}_1 = 5\times10^{-14} M_\odot h^{-2} \mathrm{Mpc}^2$ \citep{brown02,hirata04}. The implementation of all of the above has been validated in \citet{y3-methods} for DES Y3. Note that the sign convention in Eq. \ref{eq:tatt_a1} and  \ref{eq:tatt_a2} ensures consistency at the level of the power spectrum contributions. That is, if $A_1$ and $A_2$ have the same sign, then the power spectrum contributions in Eq. \ref{eq:tatt_gi}-\ref{eq:tatt_ii_bb} will do too (see \citealt{blazek19} for discussion). 

We consider a few different IA setups in this work. The most complex is the TATT model with five free parameters $(A_1,A_2,\eta_1,\eta_2,b_{\rm TA})$, which are varied with the priors shown in Table \ref{tab:params}. Alternatively the NLA model is a subspace of TATT with $A_2,\eta_2,b_{\rm TA}=0$ (i.e. 2 free parameters). In this case, both GI and II spectra have the same shape as the matter power spectrum $P_\delta$, modulated by the amplitude in Eq. \eqref{eq:tatt_a1}. Note that this is a specific variant of the NLA model introduced by \citet{joachimi11}; the version first proposed in \citet{hirata04,bridle07,hirata07} did not have the extra freedom in redshift, and had only one free parameter, $A_1$ (equivalent to fixing $\eta_1=0$ in Eq. \eqref{eq:tatt_a1} above). Where relevant, we will refer to the simpler version with only $A_1$ free, as ``1-parameter NLA" or NLA-1.

\subsection{Free parameters, sampling and priors}\label{sec:theory:priors}

All likelihood analyses used in this work are carried out within the framework of \blockfont{CosmoSIS}\footnote{\url{https://cosmosis.readthedocs.io/en/latest/}} \citep{zuntz15}. Our main results make use of the \blockfont{PolyChord} \citep{handley15} nested sampling algorithm\footnote{500 live points, $\mathrm{num\_repeats} = 30$, $\mathrm{tolerance} = 0.01$} (see \citealt{campos22} Appendix D for a comparison with the \blockfont{MultiNest} sampling algorithm for our Y3 setup). When discussing best fit values in the following sections we use $10\times$ oversampled chains generated by \blockfont{PolyChord}, in the same way as \citet*{y3-cosmicshear2} and \citet{deskids23}. The core idea here is to save extra samples in parameter space, between the ones included in the final \blockfont{PolyChord} chain. This increases the density of samples and so reduces sampling noise, effectively being equivalent to running a likelihood maximiser. When calculating degrees of freedom (e.g. for $p-$values), we use an estimate for the effective number of parameters $N_{\rm par, eff}$ rather than the naive value. This is given by the expression $N_{\rm par, eff} = N_{\rm par, nai} - \mathrm{Tr}[C_\Pi^{-1}C_{\rm p}]$, where $C_\Pi$ and $C_{\rm p}$ are the covariance matrices of the prior and posterior distributions respectively (see footnote 14 of \citealt*{y3-cosmicshear2} and \citealt{raveri19}).

For a subset of chains (the LSST Y1-like ones), instead of \blockfont{PolyChord} we opt to use a sampler called \blockfont{Nautilus}\footnote{\url{https://nautilus-sampler.readthedocs.io/en/stable/}} \citep{lange23}. This choice was primarily driven by speed -- \blockfont{Nautilus} uses neural networks to find an efficient way of choosing the boundary around each set of live points, and as such is relatively fast. It is seen to produce comparably accurate posteriors to \blockfont{PolyChord}. See Appendix \ref{app:lsst_setup} for further discussion.

In addition to either six or seven free cosmological parameters $(\as,\ns,\omegam,\omegab,h,\Omega_\nu h^2$, plus $w$ in a subset of chains run in \wcdm) and up to five free parameters for IAs (see Section \ref{sec:theory:ias} above), we have a selection of nuisance parameters. These are designed to account for uncertainties related to the data and measurements, and so differ between $2\times2$pt and $1\times2$pt analyses. For the cosmic shear part of the data vector, we allow one free multiplicative shear calibration factor $m$ per redshift bin. These parameters enter simply as $\xi^{ij}_{\pm} \rightarrow (1+m^i)(1+m^j)\xi^{ij}_\pm$. Since galaxy-galaxy lensing measurements rely on the same (potentially imperfectly calibrated) shape catalogue, the same parameters also enter the $2\times2$pt data as $\gamma^{ij}_t \rightarrow (1+m^j)\gamma^{ij}_t$. The shear catalogues also have associated redshift uncertainties. We thus allow one free shift parameter $\Delta z_{\rm s}$ per bin, which translates the redshift distribution entering the equations above as $n^i_{\rm s}(z) \rightarrow n^i_{\rm s} (z- \Delta z_{\rm s}^i)$. This parameterisation, though simple, is thought to be sufficient for cosmic shear in the current generation of surveys \citep{y3-kp}.

For the lens sample, we similarly have various nuisance parameters. First of all, each lens bin has an independent bias factor $b_1^i$, which is marginalised with wide flat priors $b_1 \in [0.8,3]$. Note that for most of this work we will only consider linear bias both for generating and analysing data vectors. The exception is Section \ref{sec:results:impact_of_bias_model}, where we consider the impact of using a more complex model on our findings. For this alternative setup we have two parameters per bin, which are varied with flat priors $b_1^i \sigma_8 \in [0.67,3.0]$ and $b_2^i \sigma_8^2 \in [-4.2,4.2]$ respectively.
Each lens bin also has a shift $\Delta z_{\rm l}$, applied in the same way as with the source $n(z)$, and a stretch factor $\sigma z_{\rm l}$, which enters as:
\begin{equation}
n^i_{\rm l}(z) \rightarrow \frac{1}{\sigma z^i_{\rm l}}  
n^i_{\rm l} \left( \frac{ z - \left \langle z \right \rangle_i}{\sigma z^i_{\rm l}} + \left \langle z \right \rangle_i \right ) 
\end{equation}
(see e.g. \citealt{y3-2x2pt_maglim} Eq. 19). These two parameters, $\Delta z$ and $\sigma z$ allow changes in the mean and width of each lens $n(z)$ (though they cannot affect the higher order details of the shape).

It is also worth noting that all $2\times2$pt and $3\times2$pt analyses discussed in this paper include point mass marginalisation (see \citealt{maccrann20}, \citealt*{prat23} and \citealt{y3-methods} for details and discussion). In brief, the procedure analytically marginalises the impact of non-local lensing contributions to $\gamma_t$. This does not require extra parameters, but effectively removes information coming from very small scales. 

We consider a few different analysis setups, which we will describe in the next section. The choice of lens and source sample determines the priors on these extra nuisance parameters. We refer the reader to Section \ref{sec:data:ggl} and Table \ref{tab:params} for a summary.

\vspace{0.5cm}
\section{Synthetic Data \& IA Model Error}\label{sec:data}

\begin{figure}
    \centering
    \includegraphics[width=\columnwidth]{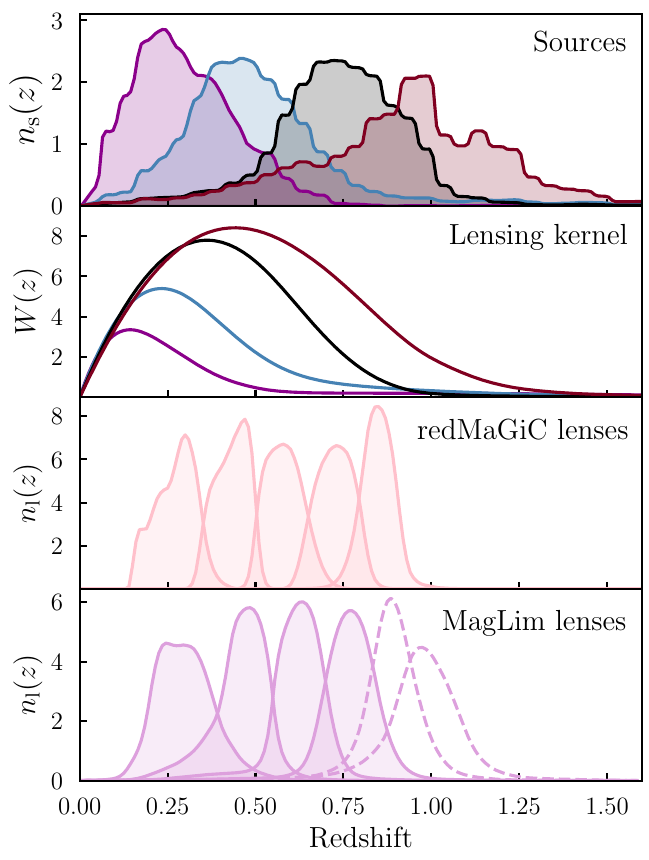}
    \caption{The fiducial redshift distributions used in this work. From top, we have (a) the DES Y3 source sample in 4 bins; (b) the associated lensing kernels, (c) the Y3 \blockfont{redMaGiC} lens $n(z)$ in five bins and (d) the equivalent distributions, but for the \blockfont{MagLim} lens sample. Note that the fiducial DES Y3 setup \citep{y3-kp} made use of the lower four \blockfont{MagLim} bins only; in this work we consider both 4- and 6-bin setups. Each of these distributions, and the overlap between them, enters the mock data as described in Section \ref{sec:theory}. For reference, the mock LSST Y1 like $n(z)$, used in Section \ref{sec:results:lsst}, are shown in Appendix \ref{app:lsst_setup}. \vspace{0.5cm}}
    \label{fig:nofz}
\end{figure}

We make use of much the same infrastructure as in \citet{campos22}. Our mock data are created with the same flat \lcdm~cosmology, $\omegam=0.29$, $A_s=2.38\times10^{-9}$, $\omegab=0.052$, $h=0.75$, $n_s=0.99$, $\Omega_\nu h^2 = 0.00083$, $w=-1$. This corresponds to $\sigma_8 = 0.79$, $S_8=0.77$ and $\sum m_\nu c^2 = 0.077$ eV. When considering DES Y3 like data, we also use the same analytic non-Gaussian estimate for the joint $\xi_\pm + \gamma_t + w$ covariance matrix, as described in \citet{y3-covariance} and used in \citet{y3-kp}. This was estimated using \blockfont{CosmoCov}\footnote{\url{https://github.com/CosmoLike/CosmoCov}} \citep{fang20}, which uses a halo model to estimate covariances, including connected non-Gaussian and super-sample variance terms, and also includes the impact of the Y3 survey mask.

\vspace{0.4cm}
\subsection{Mock cosmic shear data}\label{sec:data:shear}

We create noiseless mock data simply by running the theory pipeline with a particular set of input parameter values. As in \citet{campos22}, for our fiducial setup we use the DES Y3 source redshift distributions, shown in the upper panel of Figure \ref{fig:nofz}. We create an ensemble of 21 data vectors, each one with the same input cosmology, but different IA scenarios. A given IA scenario is defined by a set of TATT parameters (5 values per scenario), chosen using the process set out in \citet{campos22} Section 3.2, which uses a Latin hypercube to generate an initial set of samples, before rotating them to capture correlations seen in real DES Y1 data (see also their Figure 1). Note that this process is not restricted by any priors, and so we include a small number of cases outside the limits shown in Table \ref{tab:params}.

\subsection{Mock galaxy-galaxy lensing \& galaxy clustering data}\label{sec:data:ggl}

We also generate mock $\gamma_t$ and $w(\theta)$ data in the same way as described above. One complication here is the choice of lens sample; whereas in DES Y3 we had only one source catalogue, we had two separate lens catalogues. The fiducial setup made use of a magnitude limited sample (\blockfont{MagLim}), as set out in \citet{porredon21}. The idea here was to optimise the constraining power of the joint lensing and clustering data by trading redshift quality for density. Analyses using \blockfont{MagLim} lenses have slightly more nuisance parameters related to redshift error, but smaller statistical uncertainties. Due to residual systematics, the fiducial Y3 cosmology analyses made use of the first four \blockfont{MagLim} bins only \citep{y3-2x2pt_maglim}. As discussed below, however, we will consider both 4- and 6-bin \blockfont{MagLim} configurations in this work.
In addition to \blockfont{MagLim}, we consider a DES Y3 like \blockfont{redMaGiC} sample, as described in \citet{y3-2x2pt_redmagic}. Our \blockfont{redMaGiC} sample has five bins up to $z\sim0.9$, with some slight overlap between neighbouring bins. 

The redshift distributions $n_\mathrm{l}(z)$ for both our \blockfont{MagLim} and \blockfont{redMaGiC} lens samples are shown in the lower panels of Figure \ref{fig:nofz}. Although it would be simpler to consider one lens sample only, the reason we include two is because sensitivity to IAs is a function both of the shape of the lens distributions and the associated redshift uncertainties. It is, then, useful to consider a range of plausible lens samples.
Note that, as in \citet{y3-kp} but unlike some analyses in the literature, we include all possible lens-source pairs in $\gamma_t$. That is, we do not exclude bin pairs for which the mean lens redshift is higher than that of the sources. This can help to constrain IAs as well as photometric redshift error, since the lensing signal in these bin pairs is expected to be small.

\subsection{Combining the data: analysis setups}\label{sec:data:3x2pt}

All analyses in this work assume a flat cosmology with massive neutrinos. In total, our fiducial \lcdm~cosmological model has six free parameters: \omegam, \as, \ns, $h$, \omegab, $\Omega_\nu h^2$. In Section \ref{sec:results} we also run \wcdm~analyses, which have an extra free parameter, $w$, corresponding to the dark energy equation of state. One can find a summary of our priors in Table \ref{tab:params}. We constrain these parameters alongside other probe-specific ones using combinations of data, as described below. 

\begin{itemize}
\item \textbf{Cosmic shear} ($1\times2$pt): Our fiducial shear setup does not make use of any information from galaxy-galaxy lensing. We apply the fiducial scale cuts of \citet*{y3-cosmicshear2} and \citet{y3-cosmicshear1} to the $\xi_+$ + $\xi_-$ data vector, which are driven by uncertainty due to the effect of baryonic feedback on small scales. In total we have up to 19 free parameters in our fiducial setup: 6 for cosmology in \lcdm, 4 for shear calibration $m^i$, 4 for redshift error $\Delta z^i$, and either 2 (NLA) or 5 (TATT) for intrinsic alignments. 
\item \textbf{$\gamma_t + w$} ($2\times2$pt), \textbf{\blockfont{MagLim} lenses, 4 bins} : In this setup, we combine galaxy-galaxy lensing and galaxy clustering using the lowest four \blockfont{MagLim} bins as lenses. This was the fiducial choice of \citet{y3-kp}, with cuts removing the upper two bins, a choice motivated by unacceptably poor fits in \lcdm~ when those bins were included (see \citealt{y3-2x2pt_maglim} Section VII~A and Appendix~B for discussion). Again, the scale cuts are not changed from those used in the fiducial Y3 analysis. The cuts on the $2\times2$pt data correspond to lower limits on comoving separation at 6 $\mathrm{Mpc}/h$ and 8 $\mathrm{Mpc}/h$ for $\gamma_t$ and $w$ respectively (see \citealt{y3-2x2pt_maglim} Section VI-A). In this setup, we have the same parameters for shear-related systematics and cosmology as above, plus another 4 galaxy bias parameters (one per lens bin), and 8 lens redshift parameters (1 stretch and 1 shift per bin).  This gives a total of 31 free parameters for an analysis using TATT, and 28 for NLA.  
\item \textbf{$\gamma_t + w$} ($2\times2$pt), \textbf{\blockfont{MagLim} lenses, 6 bins} : This setup is the same as the one described above, but we now include the upper two lens bins. This is our most constraining and our most optimistic setup, since it uses the \blockfont{MagLim} lens sample in a regime where it was not possible in the fiducial Y3 $3\times2$pt analysis of \citealt{y3-kp} (although the upper bins have been successfully included in some variants of analyses based on CMB cross-correlations; see e.g. \citealt{chang22} Figure 8). Incorporating the extra lens bins adds another 6 parameters (2 bias + 4 redshift), making a total of 37/34 (TATT/NLA). 
\item \textbf{$\gamma_t + w$} ($2\times2$pt), \textbf{\blockfont{redMaGiC} lenses}: This analysis setup uses a mock lens sample generated with the properties of the DES Y3 \blockfont{redMaGiC} catalogue in five bins. The key difference compared with \blockfont{MagLim} is that \blockfont{redMaGiC} is designed to optimise redshift quality alone, and so the priors are slightly different (see Table \ref{tab:params} and \citealt{y3-2x2pt_redmagic}). In particular, the first four width parameters are fixed. In total, then, we have 11 lens-related parameters, and so 30/27 parameters overall. Note that this setup matches the main Y3 \blockfont{redMaGiC} selection discussed in \citet{y3-2x2pt_redmagic}. This is slightly different from the ``broad-$\chi^2$" \blockfont{redMaGiC} sample, which is also discussed in that paper. We discuss the distinction briefly in Section \ref{sec:results:impact_of_lens_sample}. 
\item \textbf{$\xi_\pm + \gamma_t + w$} ($3\times2$pt): In addition to the analyses detailed above, we also consider $3\times2$pt analyses using each lens sample.
This does not increase the number of parameters relative to the corresponding $2\times2$pt analysis, or change any of the other analysis choices, but it does significantly increase the constraining power, as we will see in Section \ref{sec:results}. 
\end{itemize}

\noindent
Finally, in Section \ref{sec:results:impact_of_bias_model} and Appendix \ref{app:sr} only we consider $1\times2$pt and $2\times2$pt analysis setups that include shear ratios. The mock shear ratio data are constructed using our simulated $\gamma_t$ measurements on small scales. Essentially shear ratios add nine data points to the data vector (the lower 3 lens bins, each with 3 different combinations of source bin pairs; see \citealt*{y3-sr} for details). When modelling these data, the nuisance parameters affecting the relevant lens and source bins are propagated through consistently.

\begin{table}
	\centering	 
	\vspace{-0.2cm}
	\begin{tabular}{ccc}
		\hline
		Parameter & Fiducial & Prior   \\ \hline
		\multicolumn{3}{c}{\textbf{Cosmology}}  \\
		$\omegam$ &  0.3 &[0.1, 0.9] \\ 
		$A_\mathrm{s}\times10^{9}$ & 2.19 & [$0.5$, $5.0$]  \\ 
		$n_{\rm s}$ & 0.97 & [0.87, 1.07]  \\
		$\omegab$ & 0.048 &[0.03, 0.07]  \\
		$h$  & 0.69  &[0.55, 0.91]   \\
		$\Omega_\nu h^2 \times 10^3$ & 0.83 & [0.6, 6.44] \\
            $w$ &  -1.0  & Fixed or $[-2,-0.333]$
		\\\hline

  		\multicolumn{3}{c}{{\bf
				  Intrinsic alignment$^{*}$}}  \\
		$A_{1}, \eta_{1}$    & 0.85, 0.0  &  [$-5,5$ ]\\
		$A_{2},\eta_{2}$ & -3.08, 2.79  & [$-5,5$ ] or $\delta(0)$\\
		$b_{\mathrm{TA}}$   & 0.12  & [$0,2$] or $\delta(0)$ \\
		\hline
		\multicolumn{3}{c}{{\bf Source photo$-z$ uncertainty}}  \\
		$\Delta z^1_{\rm s}$  & 0.0  & $\mathcal{N}(0.0, 0.018$) \\
		$\Delta z^2_{\rm s}$  & 0.0  & $\mathcal{N}(0.0, 0.013$) \\
		$\Delta z^3_{\rm s}$  & 0.0  & $\mathcal{N}(0.0, 0.006$) \\
		$\Delta z^4_{\rm s}$  & 0.0  & $\mathcal{N}(0.0, 0.013$) \\
		\hline
		\multicolumn{3}{c}{{\bf Shear calibration uncertainty}}  \\
		$m^1$ & 0.0  & $\mathcal{N}(0.0, 0.008$)\\
		$m^2$ & 0.0  & $\mathcal{N}(0.0, 0.013$)\\
		$m^3$ & 0.0  & $\mathcal{N}(0.0, 0.009$)\\
		$m^4$ & 0.0  & $\mathcal{N}(0.0, 0.012$)\\
		\hline

		\multicolumn{3}{c}{\textbf{Galaxy bias  } } 	 \\
		$b_{\rm MagLim}^{i}$  & $1.5, 1.8, 1.8, 1.9, 2.3, 2.3$ & [0.8,3.0]\\
            $b_{\rm redMaGiC}^{i}$  & $1.7, 1.7, 1.7, 2.0, 2.0$ & [0.8,3.0]\vspace{0.05cm}\\\hline
		
		\multicolumn{3}{c}{\textbf{Lens
				magnification } }   \\
		$C_{\rm MagLim}^{i} $ & 0.43, 0.30, 1.75, 1.94, 1.56, 2.96 & Fixed\\
            $C_{\rm redMaGiC}^{i} $ & $1.31, -0.52, 0.34, 2.25, 1.97$ & Fixed \vspace{0.05cm}\\ \hline

		\multicolumn{3}{c}{\textbf{Lens photo$-z$ uncertainty (MagLim) }}	  \\
		$\Delta z^1_{\rm l}$ & 0.0 & $\mathcal{N}(
		0.0, 0.007$)\\ $\Delta z^2_{\rm l}$ & 0.0 & $\mathcal{N}(
		0.0, 0.011$)\\ $\Delta z^3_{\rm l}$ & 0.0 & $\mathcal{N}(
		0.0, 0.006$)\\ $\Delta z^4_{\rm l}$ & 0.0 & $\mathcal{N}(
		0.0, 0.006$)\\ $\Delta z^5_{\rm l}$ & 0.0 & $\mathcal{N}(
		0.0, 0.007$)\\ $\Delta z^6_{\rm l}$ & 0.0 & $\mathcal{N}(
		0.0, 0.008$)\\ $\sigma z^1_{\rm l}$ & 1.0 & $\mathcal{N}(
		1.0, 0.062$)\\ $\sigma z^2_{\rm l}$ & 1.0 & $\mathcal{N}(
		1.0, 0.093$)\\ $\sigma z^3_{\rm l}$ & 1.0 & $\mathcal{N}(
		1.0, 0.054$)\\ $\sigma z^4_{\rm l}$ & 1.0 &
		$\mathcal{N}(1.0, 0.051$)\\ $\sigma z^5_{\rm l}$ & 1.0 &
		$\mathcal{N}(1.0, 0.067$)\\ $\sigma z^6_{\rm l}$ & 1.0 &
		$\mathcal{N}(1.0, 0.073$) \\\hline 

            \multicolumn{3}{c}{\textbf{Lens photo$-z$ uncertainty (redMaGiC) }}	  \\
		$\Delta z^1_{\rm l}$ & 0.0 & $\mathcal{N}(
		0.0, 0.004$)\\ $\Delta z^2_{\rm l}$ & 0.0 & $\mathcal{N}(
		0.0, 0.003$)\\ $\Delta z^3_{\rm l}$ & 0.0 & $\mathcal{N}(
		0.0, 0.003$)\\ $\Delta z^4_{\rm l}$ & 0.0 & $\mathcal{N}(
		0.0, 0.005$)\\ $\Delta z^5_{\rm l}$ & 0.0 & $\mathcal{N}(
		0.0, 0.010$)\\ $\sigma z^1_{\rm l}$ & 1.0 & Fixed \\ $\sigma z^2_{\rm l}$ & 1.0 & Fixed \\ $\sigma z^3_{\rm l}$ & 1.0 & Fixed \\ $\sigma z^4_{\rm l}$ & 1.0 &
		Fixed \\ $\sigma z^5_{\rm l}$ & 1.0 &
		$\mathcal{N}(1.0, 0.054$)\\ \hline 

	\end{tabular}
 \caption{A summary of the parameters used in the various analyses considered in this work. Here we show both fiducial (input) values used to generated data vectors and also the priors used when analysing the mock data. Square brackets $[a,b]$ indicate a flat prior within the bounds $a$ and $b$, while $\mathcal{N}(\mu,\sigma)$ is a Gaussian prior with mean $\mu$ and width $\sigma$. The upper four sections (cosmology, IAs, source redshift error and shear calibration, relate to the lensing source sample, and so enter both $1\times2$pt and all $2\times2$pt in exactly the same way. The intrinsic alignments section is marked with a star because various different input values are used in this work. We show two sets of priors in this section, for NLA and TATT setups. For the other sections, we show both \blockfont{MagLim} and \blockfont{redMaGiC} values.}\label{tab:params}
\end{table}

\subsection{Rubin LSST Year 1 Setup}\label{sec:data:lsst}

Although our primary results are based on mock DES Y3 like data, we also run a subset of our analyses using a Rubin Legacy Survey of Space and Time Year 1 (LSST Y1; \citealt{ivezic19}) like setup. A major difference here is the covariance matrix, which we recompute using \blockfont{CosmoCov} (including non-Gaussian contributions). We assume a total area of 12,300 square degrees (compared with $\sim 4143$ for DES Y3) and source and lens samples with effective number densities of  $n_{\rm eff} = 11.2$ and $n_{\rm eff}=18$ galaxies per square arcmin respectively. These are each divided between five redshift bins (all of these numbers are taken from the projections in \citealt{srd18} and \citealt{fang20}). We adopt the analytic $n(z)$ suggested by the DESC Science Requirements Document (SRD; \citealt{srd18}), but note that there is a fair degree of uncertainty in the future sample selection and redshift methods. The binning and redshift distributions are described in more detail in Appendix \ref{app:lsst_setup}. We base the input galaxy bias and lens magnification values on the \blockfont{MagLim} values shown in Table \ref{tab:params}.

When analysing the mock Rubin data, we maintain much of our previous DES Y3 pipeline including priors and other modelling choices. We update the scale cuts using the \citet{y3-methods} method of comparing baryon-contaminated and uncontaminated data vectors (although we rely on a $\chi^2$ threshold alone -- since we are only attempting a simple forecast, we do not go as far as running chains on the baryonic mock data; see Appendix \ref{app:lsst_setup} for discussion and further details). This setup is clearly an approximation. In reality, the priors in any future Rubin analysis will (hopefully) be more informative and the approach to scale cuts will likely differ from DES Y3. For the sake of simplicity, however, we do not seek to anticipate these changes. There is a strong element of unpredictability here; we cannot, for example, say how advances in either modelling of, or observational constraints on, baryonic feedback might change the eventual LSST Y1 scale cuts. Even at the precision of current data sets there are different approaches to this question. Likewise, improved understanding and control of systematics will hopefully allow tighter priors on the redshift and shear calibration (as well as potentially IA) parameters. Exactly how much tighter, however, is not something we can predict with any degree of confidence, and so we will not attempt to. We reiterate, however, that we do not need to predict all the elements perfectly in order to test the effects we are interested in here. 

\vspace{0.5cm}
\section{Results}\label{sec:results}

In this section we set out our results from analysing the 21 IA scenarios discussed in Section \ref{sec:data}. 
We begin in Section \ref{sec:results:sample2} using a single IA scenario as an example, to illustrate a core finding of this paper. In Section \ref{sec:results:dependence_on_stuff} we consider the robustness of our results to plausible changes in the analysis setup and data. We explore the generality of our results using the collection of 21 IA samples in Section \ref{sec:results:applications}. Section \ref{sec:results:mechanisms} then draws together some of the evidence to try to build a coherent understanding of why the data behave as they do. Finally we demonstrate the robustness of our findings in an LSST-like analysis setup in Section \ref{sec:results:lsst}.   

\subsection{Internal tension as an indicator of bias -- a high bias example}\label{sec:results:sample2}

\begin{figure}
    \centering
    \includegraphics[width=\columnwidth]{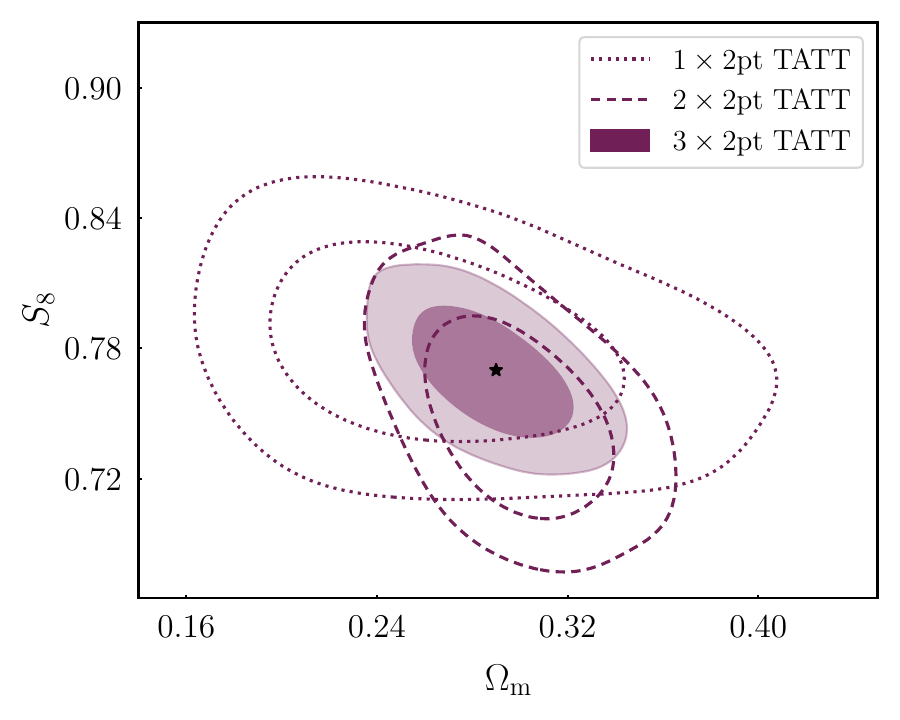}
    \caption{An example of a joint probes analysis in which the IA model (TATT) is sufficient to describe the data. Inner and outer contours represent $68\%$ and $95\%$ confidence levels. We show cosmic shear only ($1\times2$pt; dotted line), galaxy-galaxy lensing + galaxy clustering ($2\times2$pt; dashed line) and the combination ($3\times2$pt; shaded) separately. All three analyses are run on noiseless simulated data. Although there are projection effects, all recover the input (black star) to within $1\sigma$, with the differing orientations of the contours producing a relatively tight correctly-centred $3\times2$pt constraint. The lens sample used in this example is \blockfont{redMaGiC}, but a similar picture can be seen when using either \blockfont{MagLim} setup.\vspace{0.5cm}}
    \label{fig:contours_tatt}
\end{figure}

\begin{figure}
    \centering
    \includegraphics[width=\columnwidth]{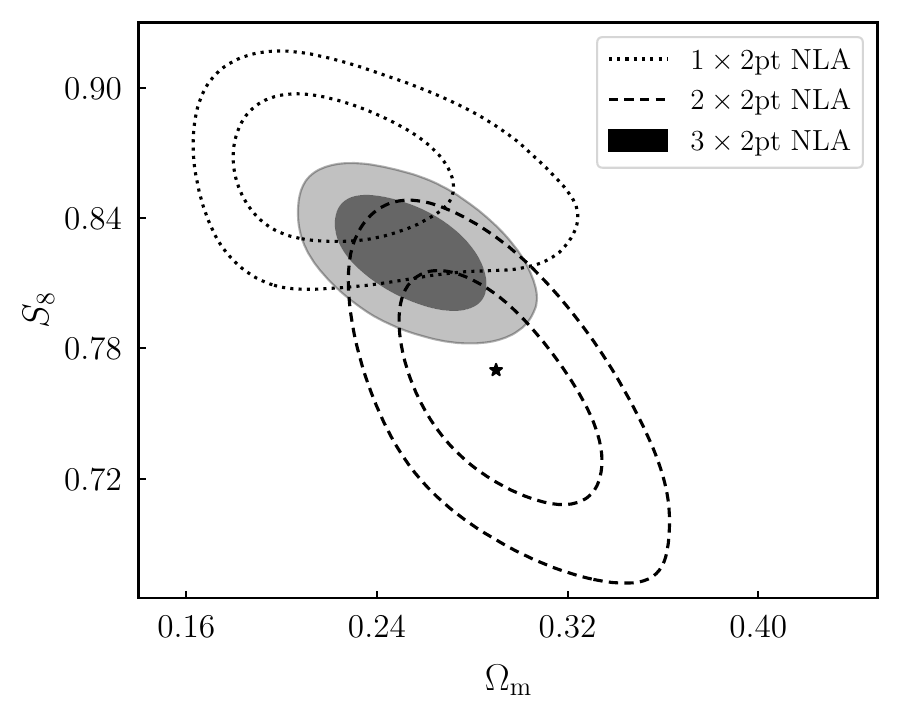}
    \includegraphics[width=\columnwidth]{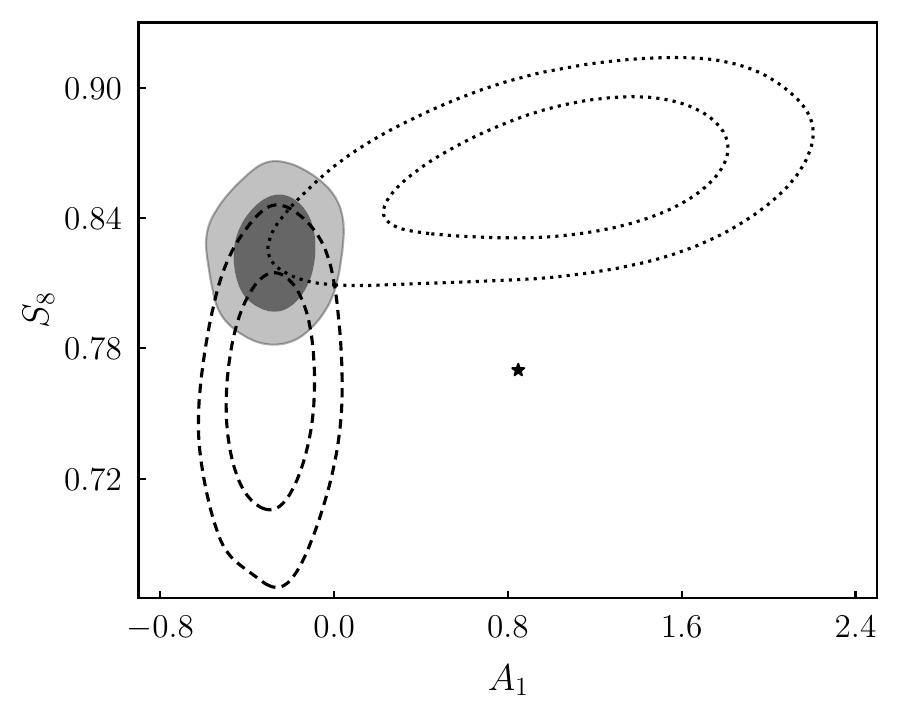}
    \caption{The same as Figure \ref{fig:contours_tatt}, but now using the 2-parameter NLA model instead of TATT. All other aspects of the analysis are the same. The top panel shows the projected $S_8-\omegam$ plane. While the cosmic shear only ($1\times2$pt; dotted) results are offset from the input by several $\sigma$, $2\times2$pt (dashed) is relatively unbiased. The two probes are discordant and the joint $3\times2$pt contours (shaded black) sit in between the two. The lower panel shows the NLA amplitude $A_1$ and $S_8$. Whereas shear predominantly absorbs the unmodelled IA signal into $S_8$, in the $2\times2$pt case it is $A_1$ that shifts.\vspace{0.5cm}
    }
    \label{fig:contours_main}
\end{figure}

For illustrative purposes, we choose a particular IA scenario to consider in more detail. This scenario has TATT parameters $A_1=0.848, A_2=-3.08, \eta_1=0.0, \eta_2=2.79, b_{\rm TA}=0.123$ (also listed in Table \ref{tab:params}). We chose this data vector from the various ones considered in \citet{campos22} because it was found to be relatively extreme, as assessed in the context of cosmic shear, resulting in a bias when using the 2-parameter NLA model of $>3\sigma$. Note that this particular parameter combination is disfavoured at $\sim 3 \sigma$ but has not been completely ruled out by the DES Y3 analyses \cite{y3-kp}. Although it was not selected for this reason, this scenario ended up being a fairly pronounced example of the differences between contamination in $1\times2$pt and $2\times2$pt cosmological constraints. As we will see in the following sections, however, the general trends hold over the range of IA scenarios.

We will start with the simplest case, where the data and the model match exactly, shown in Figure \ref{fig:contours_tatt}. For all three of the posteriors shown here, the IA model is TATT (with 5 free parameters). As we can see, both cosmic shear (dotted) and galaxy-galaxy lensing plus clustering (dashed) recover the input cosmology to $<<1\sigma$. For this example we've chosen to show the \blockfont{redMaGiC} rather than \blockfont{MagLim} 2- and $3\times2$pt results, but this choice makes no qualitative difference (we will discuss this more in Section \ref{sec:results:dependence_on_stuff}). 
As discussed in \citet{y3-methods} and \citet{y3-2x2pt_redmagic}, the $2\times2$pt case is subject to slightly stronger projection effects, offsetting the dashed contour downwards by a fraction of a $\sigma$. The constraints are entirely consistent, however, and the joint $3\times2$pt analysis recovers the input almost perfectly. The difference in contour shape nicely illustrates the point of the joint analysis -- the intersection of the two allows a relatively tight constraint on both $S_8$ and \omegam. This also demonstrates that while $S_8$ is, by design, the combination best constrained by cosmic shear, it is not quite optimal for $2\times2$pt -- there is some slope in the dashed contours in Figure \ref{fig:contours_tatt}.

We next consider the same IA scenario, but analysed with an overly simple model (2-parameter NLA). There is now a relatively strong bias in the cosmic shear fit, shown by the dotted contour in Figure \ref{fig:contours_main}. The bias here is similar in nature to that seen in the simulated tests of \citet*{y3-cosmicshear2} (see their Figure 6), with the best fitting constraints being shifted to high $S_8$ and low \omegam~by several $\sigma$. This shift is accompanied by a worsening goodness-of-fit, with $\Delta \chi^2\sim18$ for 222 degrees of freedom (227 data points; see Section 4.2.1 of \citealt{campos22}). Note that, as described in Section \ref{sec:theory:priors}, the degrees of freedom quoted here are calculated using an effective number of free parameters $N_{\rm par, eff}$, which accounts for the fact that some parameters are prior dominated, rather than the naive value from counting variables. The dashed lines in Figure \ref{fig:contours_main} show what happens when we analyse the same TATT scenario using the same over-simplified IA model, but now with a $2\times2$pt data vector. Interestingly, we can see that the bias is significantly reduced compared with the cosmic shear only case. The $2\times2$pt contour is centred on the input to well within $1\sigma$. The goodness-of-fit is slightly worse, with $\Delta \chi^2\sim50$ for $\sim297$ degrees of freedom\footnote{We have assumed the effective number of parameters in the Y3 $2\times2$pt analysis to be roughly 5, in line with cosmic shear. This is partially guided by \citet{dachuna22}, who estimated $N_{\rm p, eff}=2.51$ for a Y1 like DES $3\times2$pt analysis. Our guess accounts for the fact that Y3 is more constraining than Y1, and also has extra nuisance parameters. Although not precise, it is expected to be correct to a factor of 2 or so, and so gives us a rough idea of the quality of the fit.} (302 data points: 248 in $\gamma_t$ and 54 in $w$). In both cases, however, we can see that the change in the goodness-of-fit due to the IA contamination is relatively small compared with the width of the $\chi^2$ distribution, given the degrees of freedom. That is, \textit{under the null hypothesis that the model fits the data perfectly}, we expect a reduced $\chi^2$ of $\sim 1$, on average. If we treat the $\Delta \chi^2$ values above as coherent shifts in the goodness-of-fit, we obtain $\chi^2/N_{\rm dof} \sim (222+18)/222 = 1.08$ and $(297+50)/297\sim1.17$. In other words, despite the biases seen in Figure \ref{fig:contours_main}, it is unlikely that the goodness-of-fit would flag either the cosmic shear or galaxy-galaxy lensing + clustering analyses as problematic.

We illustrate this by rerunning both with a noisy data vector (we use the fiducial noise realisation from \citealt{campos22} here). We obtain best fits of $\chi^2_{2\times2\mathrm{pt}} = 313.3$ ($\chi^2/N_{\rm dof} \sim 1.05$, again with 222 dof) and $\chi^2_{1\times2\mathrm{pt}} = 238.2$ ($\chi^2/N_{\rm dof} \sim 1.07$, 297 dof), respectively. This gives a $p-$value $p(>\chi^2) = 0.22$ for cosmic shear and $p(>\chi^2) = 0.24$ for $2\times2$pt\footnote{Even using the naive calculation for the degrees of freedom, ignoring the impact of priors, we have $302-27=275$ degrees of freedom in the \blockfont{redMaGiC} $2\times2$pt NLA analysis and $227-16=211$ for $1\times2$pt. This still gives a $p>0.05$ in both cases.}. To summarise, there are data vector level residuals, which are apparent in the non-zero $\chi^2$ values from noiseless data vectors. Once we are in a scenario with (DES Y3 like) noise, however, it is difficult to tell anything is wrong from a theoretical interpretation of the goodness-of-fit statistics. Both $p-$values are well above commonly used thresholds such as $p=0.01$ or $p=0.05$, despite the fact that there are considerable parameter biases. This is consistent with the findings of \citet{campos22}, who reported that for cosmic shear alone, theoretically motivated thresholds based on $p-$values or information criteria are not a reliable way of identifying parameter bias (see the discussion in their Section 5.3).

It is also interesting to note that while $S_8$ (and other cosmological parameters) are largely unbiased in the $2\times2$pt analysis, there is a significant shift in $A_1$ (lower panel of Figure \ref{fig:contours_main}). While clearly it cannot do so perfectly (hence the non-zero $\Delta \chi^2$), the IA model is absorbing some of the error, resulting in a slightly negative best fit $A_1$. We will explore this trend further, and what it tells us, in Section \ref{sec:results:dependence_on_stuff}. Interestingly, we do not see significant shifts in other parameters (illustrated in Appendix \ref{app:other_params}). Although there are minor offsets in the lowest source bin, $\Delta z^1_{\rm s}$, these are well within $1\sigma$; all of the other shift and stretch redshift parameters are remarkably stable in all cases. Although it is often stated that redshift errors can easily be absorbed by IA parameters (and the other way round), and this is certainly true in the case where we have little/no prior knowledge, it is worth remembering that the redshift parameters in our case are relatively tightly controlled (see the priors in Table \ref{tab:params}). This restricts the amount of possible interaction between the two, and likely explains the relative stability that we see. A similar picture is seen with shear biases. Although galaxy bias is varied with wide priors, it is tightly constrained by $w(\theta)$, and we see no signs of significant offsets in $b^i$. Nor is there any evidence that the parameter bias in the $2\times2$pt case is simply being absorbed into other parts of cosmology parameter space; $h$, \ns, \as, \omegab~are all well centred compared with their input values. 

The joint $3\times2$pt analysis (shaded black contour in Figure \ref{fig:contours_main}) is clearly biased, but this is driven by the cosmic shear data favouring high $S_8$. That is, what we are seeing is a differential sensitivity to IA mis-modeling between galaxy-galaxy lensing and cosmic shear, but that does not translate into self-calibration in the joint analysis. Accurate modelling is still clearly needed to produce unbiased joint probe results. These results do, however, suggest that IA model error can manifest itself as a form of internal tension between different parts of the $3\times2$pt data vector. This has a number of practical implications, which we will return to later on. Note that this observation -- that cosmic shear is considerably more susceptible to parameter bias than $2\times2$pt when the IA model is wrong -- is a core finding of this paper. In the following sections we will explore what is going on in more detail. 

When considering our results, it is also worth keeping in mind that, in practice, these sorts of analyses are typically significantly correlated. That is, $\gamma_t$ and $\xi_\pm$ are measured using a common shear catalogue with the same realisation of both shape noise and cosmic variance. While one might expect to see random shifts of $1-2\sigma$ between independent data sets relatively often, the chance of this happening in a case like ours is considerably lower (see e.g. \citealt*{y3-internal_tensions} for discussion). In other words, the ``tension" as assessed using metrics such as PPD may be more significant than implied by a naive interpretation of the offset between the contours in terms of $\sigma$. 

\subsection{Dependence on analysis choices}\label{sec:results:dependence_on_stuff}

Given the observations above it is reasonable to ask how far our results extend to other analysis configurations. In this section we consider a range of variations on the fiducial setup. These tests help to confirm the generality of our main result, and feed into the discussion of the mechanism(s) producing it in Section \ref{sec:results:mechanisms}. 

\subsubsection{Dependence on the choice of lens sample}\label{sec:results:impact_of_lens_sample}

\begin{figure}
    \centering
    \includegraphics[width=\columnwidth]{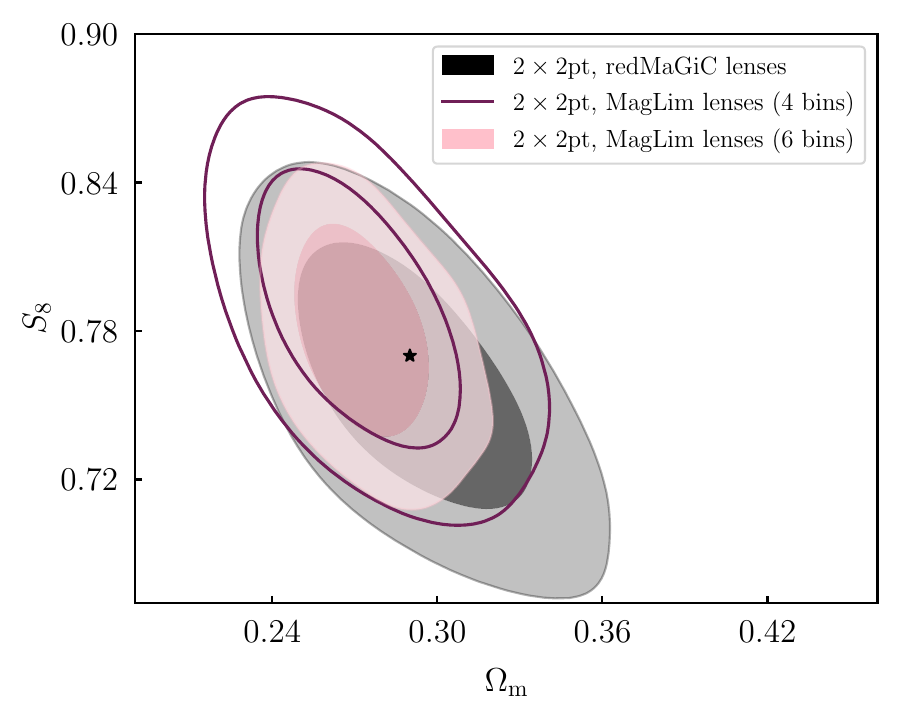}
    \caption{Simulated parameter constraints from galaxy-galaxy lensing plus galaxy clustering. Three different lens configurations are shown here (see Section \ref{sec:data:ggl} for details). In each case the input IA scenario is the same, as specified in Section \ref{sec:results}. All three lens setups recover the input cosmology (black star) to within $1\sigma$. For reference, a cosmic shear analysis with the same input IA signal is biased by $\sim4 \sigma$.\vspace{0.5cm}}
    \label{fig:contours_lenses}
\end{figure}

One plausible variation is the choice of lens sample. In particular, throughout the previous section we used the Y3 \blockfont{redMaGiC} lens sample, which has high quality redshift estimates. In a sample like \blockfont{MagLim}, the redshift uncertainties are slightly larger, requiring extra free parameters for the shapes and positions of the lens bins.

In Figure \ref{fig:contours_lenses}, we show the same IA scenario as discussed in Section \ref{sec:results:sample2}, but with a few different lens samples. The black contours are the same as in Figure \ref{fig:contours_main}, but overlain in purple and pink are the equivalent $2\times2$pt \blockfont{MagLim} results (see Section \ref{sec:data} for an explanation of the difference between the 4- and 6-bin \blockfont{MagLim} setups). As expected, these contours are tighter, for the reasons discussed in \citet{porredon21}. Although the centring here is less accurate, we can see a similar qualitative picture as before. The $2\times2$pt results are less biased than $1\times2$pt, and there is a considerable offset between $1\times2$pt and $2\times2$pt analyses on the same joint data. We consider the difference in more detail in Appendix \ref{app:lens_samples}, but note that the basic results are robust to reasonable changes in the lens sample. 

The offsets we do see in Figure \ref{fig:contours_lenses} are thought to arise from the additional redshift uncertainty in the cases with \blockfont{MagLim} lenses. The stretch and shift parameters have some freedom to increase the lens-source bin overlap in certain parts of the data vector (and, so, change the way in which IAs enter the data). We will return to this in Section \ref{sec:results:mechanisms}, but it effectively creates regions of parameter space where $S_8$ is more degenerate with the unmodelled IA signal, and so reduces the ability of the data to distinguish the two. 

In addition to the main Y3 \blockfont{redMaGiC} lens sample, \citet{y3-2x2pt_redmagic} also considered what became known as ``broad-$\chi^2$ \blockfont{redMaGiC}". It was shown in that paper that relaxing the \blockfont{redMaGiC} goodness-of-fit criterion reduces colour-dependent photometric selection biases. Since not all of the calibration steps were run on the new broad-$\chi^2$ \blockfont{redMaGiC} sample, nor has it been validated to the same extent as the main \blockfont{redMaGiC} and \blockfont{MagLim} lens samples, we do not run mock analyses in this setup. We note, however, that any broad-$\chi^2$ \blockfont{redMaGiC} $2\times2$pt analysis would most likely sit between the black and pink contours in Figure \ref{fig:contours_lenses}, both in size and position. 
In terms of number density, the broad-$\chi^2$ sample is between \blockfont{MagLim} and fiducial \blockfont{redMaGiC}, and so the $2\times2$pt constraining power is expected to also be somewhere between the two \citep{porredon21}. The predicted broad-$\chi^2$ \blockfont{redMaGiC} redshift priors are also wider than for fiducial \blockfont{redMaGiC}, but still somewhat tighter than those for \blockfont{MagLim} (see \citealt{y3-2x2pt_redmagic} Table II and Appendix C). For this reason, we would not expect to see shifts more extreme than those in Figure \ref{fig:contours_lenses}.

\subsubsection{Dependence on IA model}\label{sec:results:model_dependence}

\begin{figure}
    \centering
    \includegraphics[width=\columnwidth]{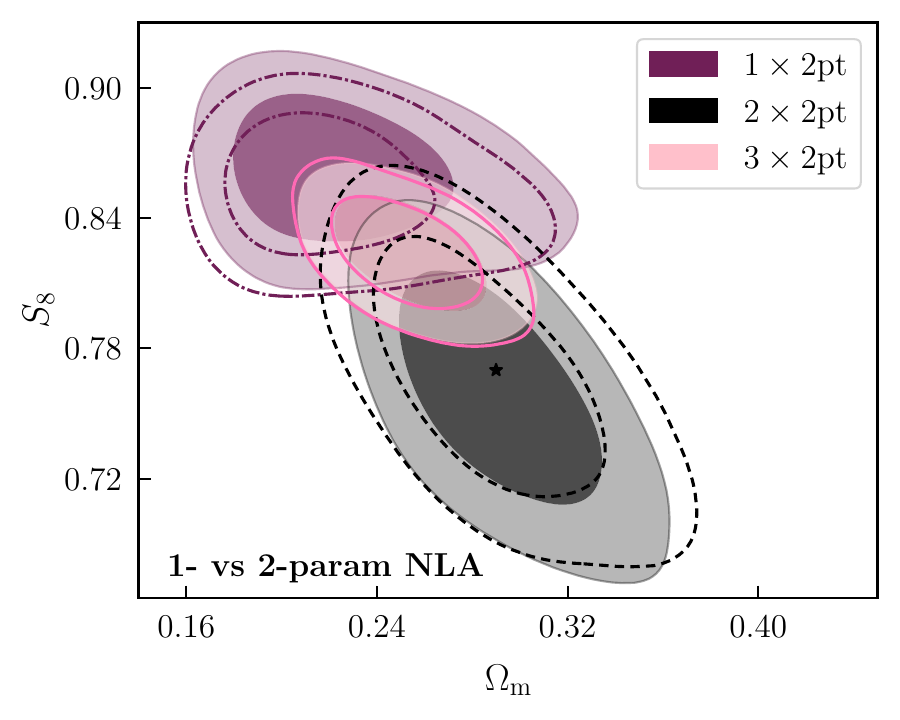}
    \includegraphics[width=\columnwidth]{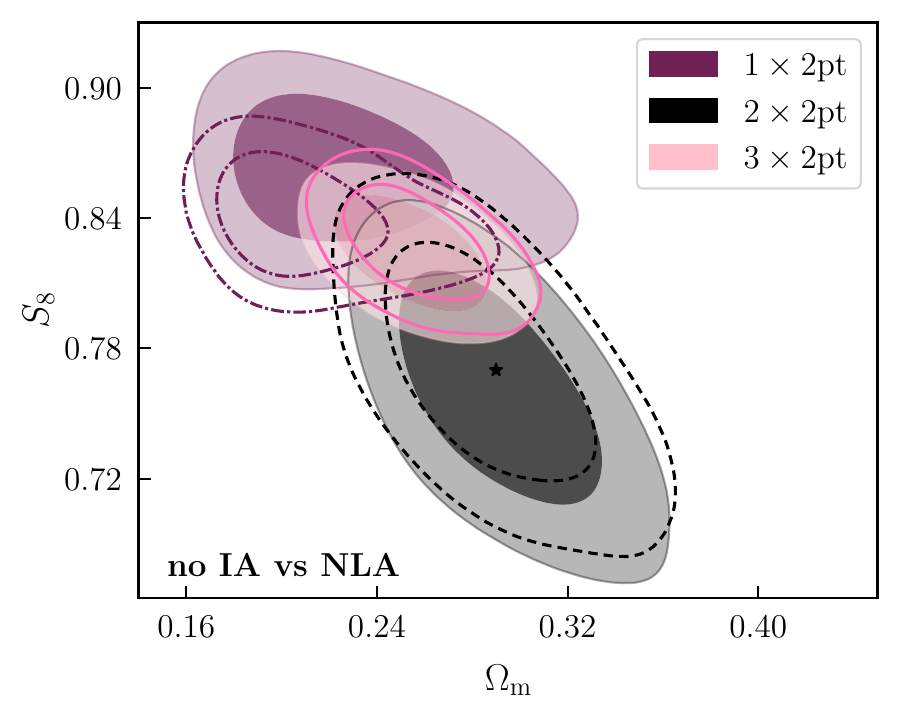}
    \caption{The impact of the choice of IA model. The shaded contours here are the same as those shown in the upper panel of Figure \ref{fig:contours_main} (NLA with 2 free parameters). Overlain (open contours) are analyses of the same three contaminated data vectors, but using an even simpler IA model (NLA with only one free parameter in the upper panel, no IA model at all in the lower). Though there are shifts, the qualitative picture is the same: galaxy-galaxy lensing plus clustering is significantly less biased than either shear alone or the full $3\times2$pt combination.\vspace{0.5cm} }
    \label{fig:ia_model}
\end{figure}

Another question one could ask is: how specific are these observations to our choice of IA model. After all, the unmodelled TATT signal in $\gamma_t$ appears to be preferentially absorbed by the NLA parameters rather than cosmology (see Figure \ref{fig:contours_main} and the discussion in Section \ref{sec:results:sample2}). It is reasonable to ask how far this is a function of the flexibility of the IA model. For our main results, we chose to analyse TATT-generated data with a 2-parameter ($A_1$, $\eta_1$) variant of the NLA model (which was the fiducial choice of \citealt{deskids23} and was also run as an analysis variant in \citealt*{y3-cosmicshear2} and \citealt{asgari21}). Now we try running a simpler sub-model, with the redshift index fixed $\eta_1=0$. This is closer to the original NLA model of \citet{hirata04}, and was used by KiDS-1000 \citep{asgari21,heymans21,troster22}. For comparison we also run a version assuming no IAs at all (i.e. $A_1,A_2=0$). This is not a realistic modelling choice, but we include it here for illustrative purposes. 

The results are summarised in Figure \ref{fig:ia_model}. Switching to NLA-1 (upper panel) causes small shifts in both the shear and $2\times2$pt contours. In the former case, the contours are already at the edge of the prior space in almost all of the cosmology parameters -- although not quite hitting the edge in the projected \omegam~direction, it \emph{is} in the 2D $\sigma_8-\omegam$ plane, as well as in $h$, $\ns$ and $\omegab$ (this can be seen in Appendix \ref{app:other_params}, which shows the wider cosmological parameter space with the prior bounds). The purple contour is thus restricted in how much further it can go in the high-$S_8$/low-\omegam~direction. The impact of switching models on $2\times2$pt is also small, but for different reasons. Here we see a very small upwards shift, with the $1\sigma$ bound still comfortably enclosing the input. If we consider the extreme model setup in the bottom panel, we see interestingly little difference. Here we are not modelling IAs at all. The $1\times2$pt contour is seen to shrink and shift downwards slightly. This, again, is likely due to the restriction of the prior preventing it moving further upwards\footnote{Note that the fan-shaped prior in the $S_8-\omegam$ plane (see Figure \ref{fig:giant_plot_cosmology}) is a result of our decision to sample $\as$ rather than $S_8$. One could conceivably widen it by expanding the \as~prior, but we note that the bounds are already considerably wider than the range allowed by Planck 2018 (at the level of 10s of $\sigma$).}. 

The $2\times2$pt case is interesting here -- unlike with cosmic shear, the constraints are well contained by the prior, and so edge effects are not a factor. In both (1- and $2\times2$pt) cases, the best $\chi^2$ gets steadily worse as one simplifies the model from 2-parameter NLA to 1-parameter NLA to zero IA. The change is still relatively modest, however, with $\Delta \chi_{2\times2\mathrm{pt}}^2 \sim 50 \rightarrow 56 \rightarrow 58$. We can conclude a few things here. First, given the small change in $\chi^2$, it may be that we are seeing some level of cancellation between IA contributions, such that the preferred $A_1$ happens to be close to zero. This can happen in galaxy-galaxy lensing in a way that is not, in general, possible with cosmic shear (although, as we will come to in Section \ref{sec:results:mechanisms}, there are reasons to think this is not the full story). Another conclusion we can draw is that while there \emph{is} residual model error, which gets worse as the model is simplified, it simply is not strongly degenerate with $S_8$ in the way it is for cosmic shear.

The net result is that, regardless of the details of the wrong IA model, internal tension can arise due to differences in how IAs enter the different probes. This is interesting as it points to the effect being relatively general, rather than specific to our Y3 setup. 

It is also notable that contrary to the basic expectation, the $2\times2$pt contour in the upper panel \emph{widens} slightly as the IA model is simplified. There are a few possible factors behind this. We should note first that for this particular input IA scenario, $\eta_1$ prefers large values, to the point where the posterior is hitting the upper edge of the $[-5,5]$ prior. Although it does not visibly distort the shape of the posteriors, this potentially reduces the size of the NLA-2 contour marginally. We should also keep in mind that the best-fit is worsened as model parameters are removed, which can change and potentially flatten the shape of the likelihood distribution, resulting in small differences in the size of the contours. One other possible factor is related to the size of the IA signal. In the NLA-1 case, the marginalised constraint on $A_1$ is relatively close to zero. For 2-parameter NLA, the posterior favours slightly larger negative $A_1$ values. Since NLA predicts a contribution proportional to the matter power spectrum $P(k)$, the signal carries information about cosmological parameters as well as IAs. As a result, scenarios in which the IA contribution is strong and well-constrained allow marginally tighter constraint on cosmology than ones without IA (see also \citealt{bridle07} Fig. 8). We should note however that this is expected to be a small effect, and the shift in $A_1$ is also small, so we do not think this is the main factor at work here.

\subsubsection{Dependence on scale cuts and galaxy bias model}\label{sec:results:impact_of_bias_model}

As discussed in Section \ref{sec:theory} the main DES Y3 analyses opted to model galaxy bias using a linear approximation, with scale cuts to facilitate this. An alternative approach would be to extend to smaller scales, at the cost of needing to marginalise over a more complex bias model. To test the impact, we re-run a subset of our $2\times2$pt chains in such a small scales + nonlinear bias setup. More specifically, we use the parameterisation of \citet{pandey22}, which has been shown to be accurate at the level of a few percent down to scales of $4$ Mpc$/h$ (see also Section \ref{sec:theory:ggl}). This setup was used on the real data, and so has been fairly extensively validated for both the \blockfont{redMaGiC} and \blockfont{MagLim} lens samples at Y3 precision \citep{y3-2x2pt_maglim,y3-2x2pt_redmagic}. As explained in Section \ref{sec:theory}, our modelling follows DES Y3, and so both our fiducial and our small scale/nonlinear bias $2\times2$pt setups include point mass marginalisation (see \citealt{y3-kp} and \citealt{y3-methods}).

\begin{figure*}
    \centering
    \includegraphics[width=\columnwidth]{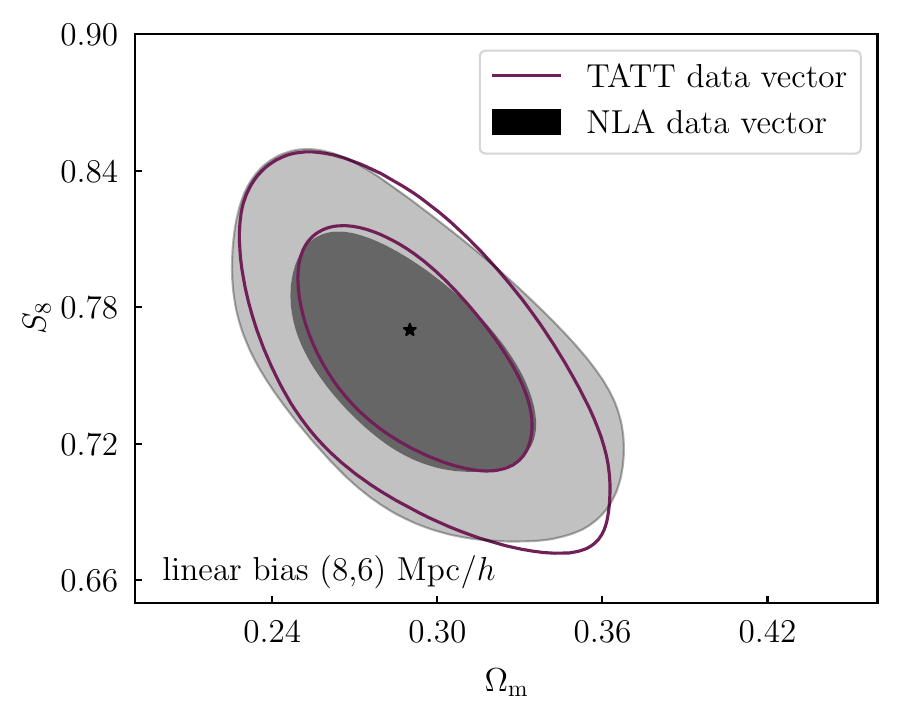}
    \includegraphics[width=\columnwidth]{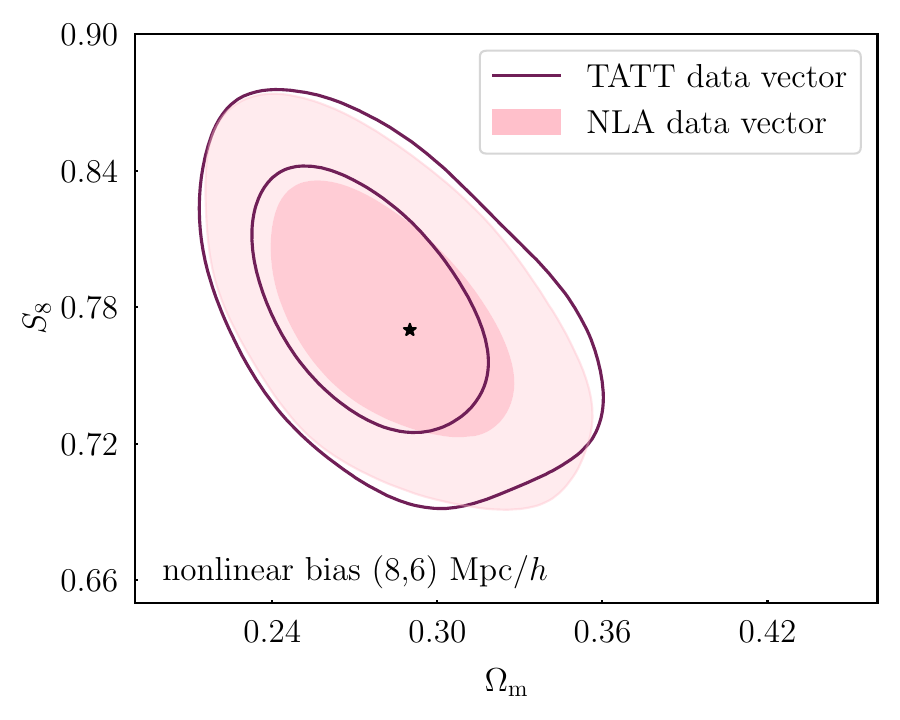}
    \includegraphics[width=\columnwidth]{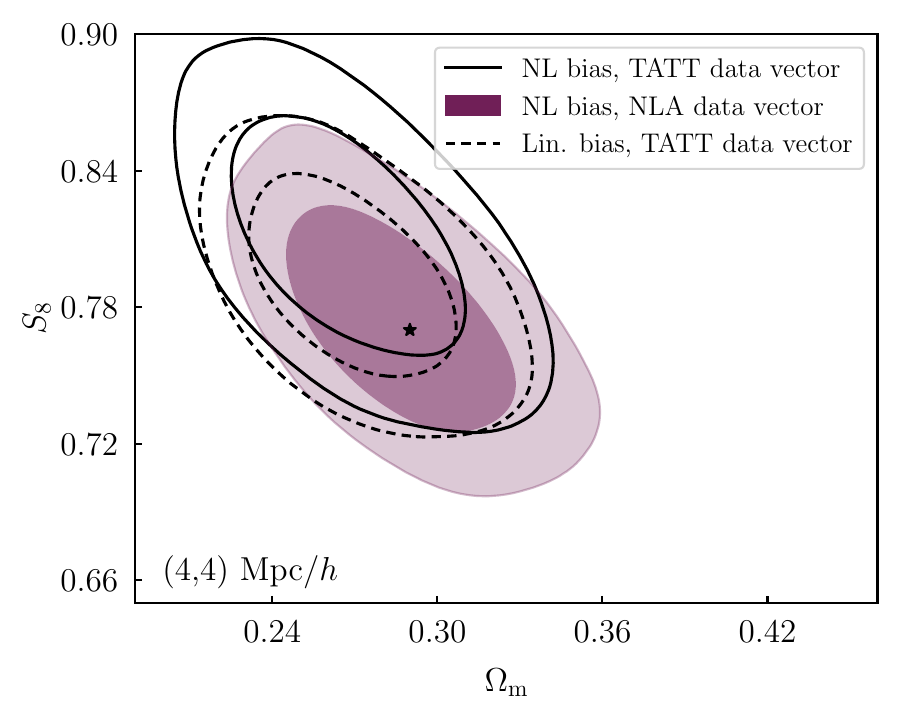}
    \caption{The impact of the galaxy bias model and scale cuts on our fiducial $2\times2$pt~analysis. We show three different combinations of bias model and scale cuts: linear bias only with large scale cuts ($>8$ Mpc$/h$ for $w$, $>6$ Mpc$/h$ for $\gamma_t$; upper left, black); nonlinear bias + large scale cuts (upper right, pink); nonlinear bias + small scale cuts ($>4$ Mpc$/h$ for both $w$ and $\gamma_t$; bottom centre, purple). In all cases shown, the IA model is 2-parameter NLA and the simulated data contain linear bias. In each case we show chains run on an ``uncontaminated" NLA data vector, as well as our fiducial TATT scenario. The difference between the two represents the bias due to IA mismodelling, excluding projection effects. As discussed in Section \ref{sec:results:impact_of_bias_model}, the lower panel also shows a case using linear bias combined with $(4,4)$Mpc$/h$ cuts to illustrate the effects of bias model flexibility.\vspace{0.5cm} }
    \label{fig:contours_bias_model}
\end{figure*}

Our findings are summarised in Figure \ref{fig:contours_bias_model}. 
To begin we repeat our baseline $2\times2$pt analysis on the TATT data vector introduced in Section \ref{sec:results:sample2}, but using the third-order bias model mentioned above. As before, these data contain an unmodelled higher order IA signal, but the input galaxy bias is linear, and there are no other sources of error. The impact can be seen by comparing the upper two panels in Figure \ref{fig:contours_bias_model}. For reference, we also show the same model run on an NLA version of the data vector. Although there is an overall shift due to projection effects, in both cases the bias due to the IA contamination is negligible (i.e. chains run on TATT and NLA data vectors agree very well). 

Next we repeat the same NLA + nonlinear bias analysis, but incorporating scales down to 4 Mpc$/h$ in both parts of the $2\times2$pt data vector. In total this increases the number of data points from 302 (248 in $\gamma_t$ + 54 in $w$) to 350 (280 + 70). This results in the black posterior in the lower panel of Figure \ref{fig:contours_bias_model}. Although still considerably less biased than the cosmic shear analysis of the same IA scenario (see Section \ref{sec:results:sample2}), there is an upwards shift of around $0.7\sigma$. Unlike in the upper panels, we cannot ascribe this to projection effects, since the chains run on NLA and TATT data vectors are also offset from each other. 

It is worth briefly considering what this tells us about the mechanisms behind the robustness of the $2\times2$pt data. The shift between the purple and black contours in the lower panel of Figure \ref{fig:contours_bias_model} could be caused by two things. One explanation is that shifting to $4$ Mpc$/h$ simply includes more data points that are strongly affected by the higher order TATT contamination; this can worsen the overall $\chi^2$, but also change the balance of bin pairs affected/unaffected by IAs. This can potentially affect the ability of the data to self-calibrate, as we will come to in Section \ref{sec:results:mechanisms}. An alternative explanation is that the extra model freedom allowed by the (now relatively well constrained) $b_2$ parameters is allowing IA error to cross over into cosmological error. The idea of degeneracy-breaking, discussed further in Section \ref{sec:results:mechanisms}, relies on the lensing dominated parts of $\gamma_t$ being able to exclude shifts in cosmology that could otherwise absorb error in the IA dominated parts. Certain kinds of extra model freedom can break (or weaken) this self-calibration by allowing the lensing dominated parts of the data to adjust. We can test the second hypothesis by rerunning the $4$ Mpc$/h$ analysis with linear bias. This is not a realistic setup, but is designed to shed light on the effects at work here. As we can see from these results (the dashed contours in the lower panel of Figure \ref{fig:contours_bias_model}), the bias model freedom alone accounts for roughly half of the bias. The rest, we conclude, is coming from the stringer unmodelled TATT contamination on smaller scales.

In addition to the above tests, we also consider the impact of including information from small scale shear ratios, in the way described in \citet*{y3-sr}. More details can be found in Appendix \ref{app:sr}, but in brief there is very little change in how $2\times2$pt responds to IA error even including shear ratio measurements down to $2$ Mpc$/h$.

When thinking about the impact of scale cuts, we should note some basic differences between our cosmic shear and $2\times2$pt data, which affect the physical scales they respond to. Our cuts for $2\times2$pt, are defined in terms of a fixed physical separation (see Section \ref{sec:data:ggl}), and translated into an angular cut in each lens bin. This puts a fairly simple lower limit on the scales on which we are sensitive to IAs. The same approach is not possible for cosmic shear, both because of the much broader redshift distributions, but also because the lensing kernel tends to mix physical scales. Our $1\times2$pt cuts are, then, defined in angular space. The result is that the physical cutoff varies between bin pairs, and a range of scales are included in the final analysis (see \citealt*{y3-cosmicshear2} Section IV-B and Figure 4). Shear analyses tend to be sensitive to physics on smaller scales, and also to have slightly less control over exactly which are used, compared with $2\times2$pt. Additionally, point mass marginalisation further removes information coming from small scales. Given all this, it makes some intuitive sense that scale cuts may be more effective in mitigating higher-order IA terms in $\gamma_t$ than $\xi_\pm$. For reasons we will come to in Section \ref{sec:results:mechanisms}, we do not think this alone can explain the robustness of $2\times2$pt analyses, but it is a factor that likely plays a role.

In summary, we do see some sensitivity to the choice of $2\times2$pt cuts, with our small-scale galaxy-galaxy lensing + clustering analysis being more susceptible to IA-related biases than our fiducial large scale one. As we have seen, however, for realistic setups the impact is still relatively small. Our result from the previous sections does not change qualitatively, with the bias in a $2\times2$pt analysis being considerably smaller than in a cosmic shear one based on the same data. We have presented tests using models that are currently thought to be robust. In coming years it is likely that there will be advances in modelling of $\gamma_t$ on small scales (see e.g. \citealt{zacharegkas22} and the discussion in that paper). In such cases we should give careful consideration to sensitivity to IA error, and how the mechanisms we have described in this paper may change. 

\subsection{Alternative input IA scenarios}\label{sec:results:applications}

\begin{figure}
    \centering
    \includegraphics[width=\columnwidth]{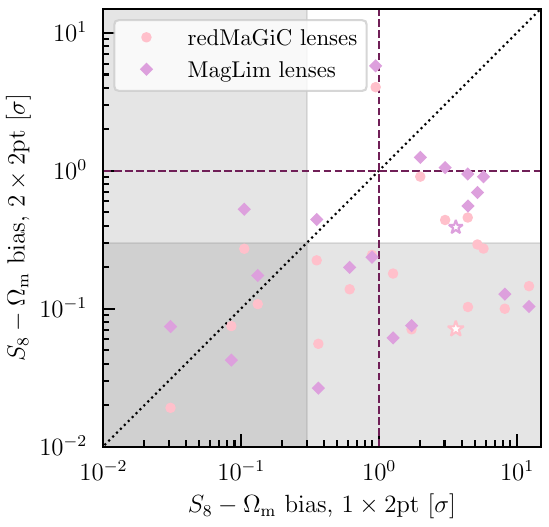}
    \caption{Bias in the 2D $S_8-\omegam$ plane in mock cosmic shear (horizontal axis) and galaxy-galaxy lensing plus galaxy clustering (vertical axis) analyses. Each point represents a particular input IA scenario, and the bias arises from an insufficient IA model (in this case 2-parameter NLA). The fiducial IA scenario, discussed in Section \ref{sec:results:sample2} is represented by the open stars. The diagonal line marks bias equality (i.e. the bias in the $2\times2$pt and $1\times2$pt analyses are exactly equal). The dashed lines represent $1\sigma$ bias, and the shaded bands show the $<0.3\sigma$ region for each probe. We show results using two alternative lens samples here, \blockfont{redMaGiC}~and \blockfont{MagLim} (pink and purple, as labelled; see Section \ref{sec:data:ggl} for details).}
    \label{fig:bias_1x2pt_2x2pt}
\end{figure}

\begin{figure}
    \centering
    \includegraphics[width=\columnwidth]{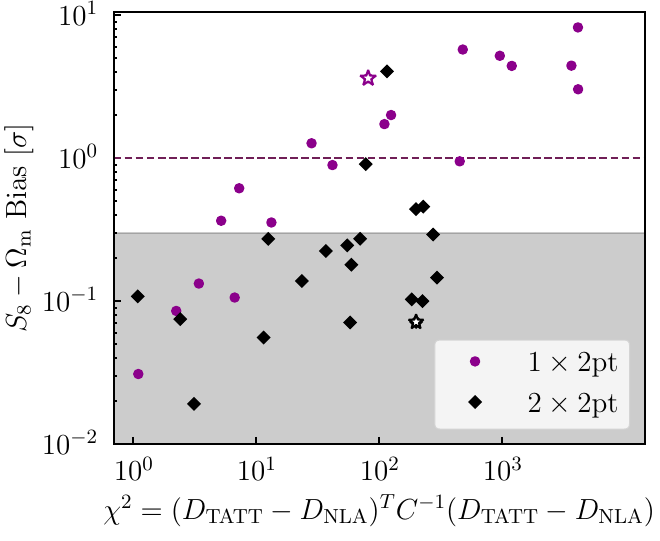}
    \caption{Data vector contamination and the resulting cosmological bias for cosmic shear (purple circles) and galaxy-galaxy lensing plus clustering (black diamonds). ``Contamination" (the x-axis quantity) is defined as a $\chi^2$, evaluated between noiseless TATT and NLA data vectors at the same fiducial input cosmology. This is effectively a summary statistic for the impact of higher-order IA contributions. In general, we can see that for a given level of contamination, the bias in the cosmic shear analyses tends to be larger than for $2\times2$pt. For clarity, we only show the \blockfont{redMaGiC}~lens sample here. As in Figure \ref{fig:bias_1x2pt_2x2pt}, the grey band and dashed line show $0.3\sigma$ and $1\sigma$ thresholds in bias. We again highlight our fiducial IA scenario as open stars.\vspace{0.5cm}}
    \label{fig:chi2_initial_bias}
\end{figure}

We next move from our single example to look at whether our conclusions bear out when considering a range of IA scenarios. Although we do not think cancellation is the driving factor in the differential $1\times2/2\times2$pt sensitivity, it is still worth testing that our fiducial IA scenario is not unrepresentative in some way. As discussed in Section \ref{sec:data}, we have a collection of 21 data vectors with different input IA scenarios, sampled from the DES Year 1 TATT posteriors. These scenarios were chosen to cover a wide range of plausible IA parameter space, including cases that are unlikely but not completely ruled out by existing data. The selection process is set out in more detail in \citet{campos22} Section 3 (see also their Figure 1).   

In Figure \ref{fig:bias_1x2pt_2x2pt} we show the relation between cosmological bias in $1\times2$pt and $2\times2$pt NLA analyses of each of these samples. Our metric for cosmological bias here is defined as an offset in the 2D $S_8-\omegam$ plane relative to the input, and is estimated as described in \citet{campos22} Section 4.1. We show 21 different input IA scenarios, with each point representing a particular scenario. For reference we indicate our fiducial case as a star. As we can see, the fiducial case is not particularly unusual. In general, $2\times2$pt is less biased than $1\times2$pt for a particular input (the points tend to lie below the diagonal dotted line). Although it is not always the case that the residual bias is completely insignificant (particularly in the \blockfont{MagLim} setup; purple diamonds), we see that $3-5\sigma$ in cosmic shear tend to translate into $1\sigma$ or less in $2\times2$pt. We should note that there is a relative rotation of the posteriors between $1\times2$pt and $2\times2$pt (i.e. $S_8$ is not the optimal combination of $\sigma_8$ and \omegam~for $2\times2$pt). This can reduce the size of a shift in the $S_8$ direction, as measured in terms of $\sigma$. Note however, that the IA induced shifts do not act purely in the $S_8$ direction -- considering the 21 scenarios, we find a fairly isotropic distribution of biases relative to the input, and there is no reason to think the contour shape will make the biases as measured in the 2D $S_8-\omegam$ plane larger or smaller. Even measured solely in the $S_8$ direction, the difference tends to be significant in absolute terms. That is to say, the trend seen in Figure \ref{fig:bias_1x2pt_2x2pt} is not thought to arise from the geometry of the contours

We should note that in the most extreme cases the cosmic shear posteriors are hitting the prior bounds in at least some parts of the parameter volume. This limits the size of the bias in absolute terms, but it also distorts the posterior, leading to an artificial reduction in the $1\sigma$ confidence region. We do not consider this an issue, since the bias in these cases is already large (i.e. we are not concerned particularly in distinguishing between $5\sigma$ and $10\sigma$ biases -- both are in the regime of unacceptably large).

Of the two lens configurations, \blockfont{MagLim} $2\times2$pt is consistently more sensitive to IA mismodelling than \blockfont{redMaGiC} $2\times2$pt (the pink points tend to be lower than the purple for a given TATT input). This, again, is consistent with what we saw previously (Section \ref{sec:results:impact_of_lens_sample} and Figure \ref{fig:contours_lenses}), and is likely due to the better redshift quality in \blockfont{redMaGiC}, as we discuss in Appendix \ref{app:lens_samples}. In both cases, however, the general conclusion that $2\times2$pt tends to be more robust than cosmic shear holds.

In Figure \ref{fig:chi2_initial_bias}, we show the data vector level contamination for the same collection of IA samples. That is, for each sample we evaluate the $\chi^2$ between the contaminated TATT data vector, and an equivalent NLA data vector (identical input parameters, but with $A_2, \eta_2, b_{\rm TA} = 0$). This serves as a summary statistic for the level of higher-order IA contamination in each simulated data vector. Note that this is not quite the same as the size of the overall IAs signal, but is the impact of the specific terms that are not captured by NLA. As we can see in Figure \ref{fig:chi2_initial_bias}, in the case of cosmic shear, there is a clear positive correlation between the amplitude of the TATT contamination and the cosmological bias. This makes intuitive sense -- as the signal NLA cannot reproduce gets larger, the bias due to mismodelling gets larger. The trend in the black ($2\times2$pt) points is interesting in comparison. Here we see much lower levels of bias for a given $\chi^2$; the black points tend to lie below the purple, on average. We should also note that there is much less (if any) correlation between bias and contamination for $2\times2$pt. Indeed, the cases with the largest bias are not the ones with the largest contamination level. This is, we should note, consistent with the hypothesis we will come to in Section \ref{sec:results:mechanisms} that the reason for the robustness of $2\times2$pt analyses is due to a lack of degeneracy with cosmology rather than inherently low levels of IA contamination in $\gamma_t$. While the distribution of black points in Figure \ref{fig:chi2_initial_bias} is noticeably lower for a given $\chi^2$, it is also true that the data-level contamination in the $2\times2$pt cases does not reach the very extreme values as $1\times2$pt (i.e. there are no black points on the far right hand side of Figure \ref{fig:chi2_initial_bias}). This is a sign that the higher order terms are having less impact, either because of internal cancellation of TATT contributions or because $\gamma_t$ (after scale cuts) is less sensitive to small scale information. Even if these are not the primary factors behind the lack of bias in $2\times2$pt, they may be acting to limit the maximum contamination. 

Considering the set of IA samples shown in Figures \ref{fig:bias_1x2pt_2x2pt} and \ref{fig:chi2_initial_bias} together, we can identify some basic trends with input parameters. First of all, the bulk of the scenarios have $A_1$ and $A_2$ with opposite signs due to the way in which they were generated. It does not appear, however, that the bias in either probe is notably smaller (or larger) in the few samples where this is not true. This implies, again, that although internal cancellation in the $2\times2$pt case may occur, it is not the primary mechanism at work.  Overall we find a fairly strong correlation between the cosmic shear bias and the low-redshift $A_2$ amplitude in the data (that is, $A_2(z) = A_2\times [(1+z)/(1+z_0)]^{\eta_2}$, evaluated at $z=0.3$). If we choose to evaluate $A_2(z)$ above the pivot redshift (i.e. $z>0.62$), the correlation weakens considerably. This suggests that, at least with current data sets, we are primarily sensitive to the behaviour of IAs at low redshift. Put another way, the IA scenarios with large positive $\eta_2$ (implying an $A_2$ signal that goes to 0 at $z \sim 0$ and increases at high $z$) are often cases in which NLA is close to unbiased. There is no obvious correlation between the same low-$z$ $A_2$ amplitude and the $2\times2$pt bias.

If we similarly consider the density weighting term $A_{1\delta}(z) = b_{\rm TA} A_1 \times [(1+z)/(1+z_0)]^{\eta_1}$, evaluated at low $z$, we see no clear correlation with the cosmic shear bias. For the bulk of samples we also see no clear trend with $2\times2$pt bias either. It is noticeable, however, the two cases with the largest $2\times2$pt bias are the two with the largest $|A_{1\delta}|$. In particular the one case with bias $>1\sigma$ (the pair of points at the top of Figure \ref{fig:bias_1x2pt_2x2pt}) has fairly small input amplitudes ($A_1=0.48$, $A_2=-0.13$), combined with very strong redshift dependence ($\eta_1=-6.6$, $\eta_2=-6.8$) and $b_{\rm TA} = 1.0$. This combination gives the largest value of $A_{1\delta} (z=0)$ of all of the samples by at least a factor of $2$. It is possible that this is a particular mode of IA error that is able to break the self-calibration mechanism in our $2\times2$pt analysis. We should also note, however, that this is an unusually extreme set of input values, with $\eta_1$ outside of our prior $\eta_1\in[-5,5]$. This means that even ignoring the higher order TATT terms, the NLA model as we have defined it cannot reproduce the data perfectly. It is somewhat difficult to draw general conclusions from this without further experimentation with more IA samples. We should note, however, that is an unusual case and it is not critical for the conclusions of this paper.

It is worth finally noting that these results rely on the assumption that any significant unmodelled IA contributions that are missed by NLA are encompassed by the 5 parameter TATT model. This is reasonable given that TATT is physically motivated, and is expected to capture the relevant processes on scales $>\sim 2$ Mpc$/h$ (see \citealt{blazek19} for discussion). We have also been careful to select a wide range of parameter values in order to explore the scope of possible impacts (see Figure 1 from \citealt{campos22}). Finally, we should note that, as we will come to in Section \ref{sec:results:mechanisms}, there is some evidence to suggest that our findings are the result of degeneracy breaking by the combination of contaminated and uncontaminated bin pairs in the $\gamma_t$ data. This is a generic effect, determined by the differential sensitivity of the data rather than the details of the unmodelled IA power spectra. Given this, it should be relatively independent of the ``complete" IA model we use for contamination.

\vspace{0.5cm}
\subsection{Possible mechanisms for the difference in sensitivity to IA error}\label{sec:results:mechanisms}

\begin{figure*}
    \centering
    \includegraphics[width=2\columnwidth]{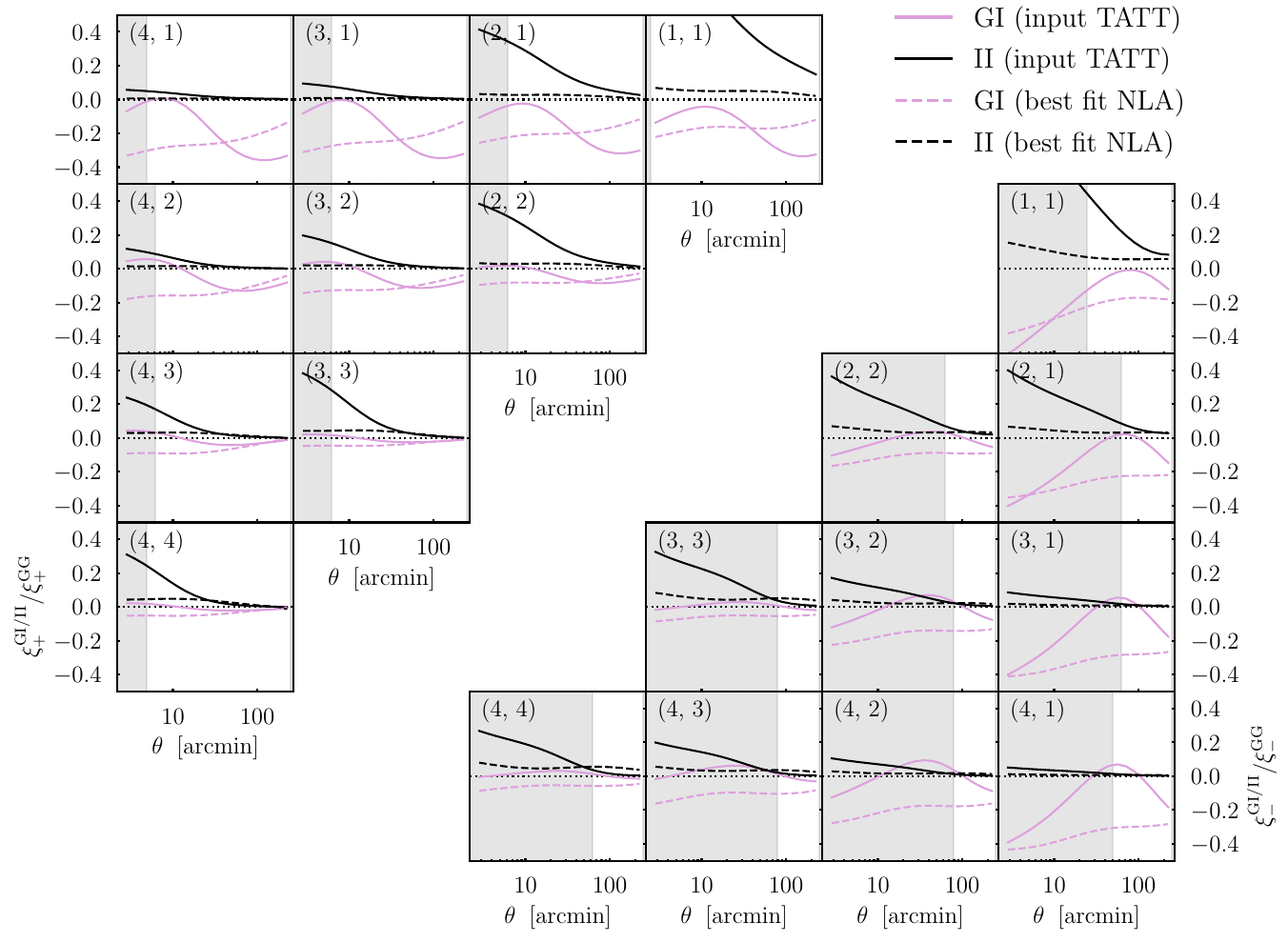}
    \caption{The fractional contributions of the GI and II IA terms to our mock cosmic shear data. Each IA component is shown as a fraction of the pure lensing signal GG, as evaluated at the input cosmology. Solid lines show the input TATT signal used to create the mock data vector discussed in Section \ref{sec:results:sample2}. The dashed lines is the IA signal as predicted at the best fit from an NLA cosmic shear analysis of the same mock data vector. Shaded grey regions indicate scales removed from our analysis (the fiducial Y3 scale cuts). We see a relatively strong IA signal, at $10-20\%$ or more on intermediate scales in most bins. Note that although we show our fiducial IA scenario here for illustrative purposes, a similar pattern can be seen when plotting the other cases discussed in Section \ref{sec:results:applications}.\vspace{0.5cm}}
    \label{fig:xipm_GI_II}
\end{figure*}

\begin{figure*}
    \centering
    \includegraphics[width=1.9\columnwidth]{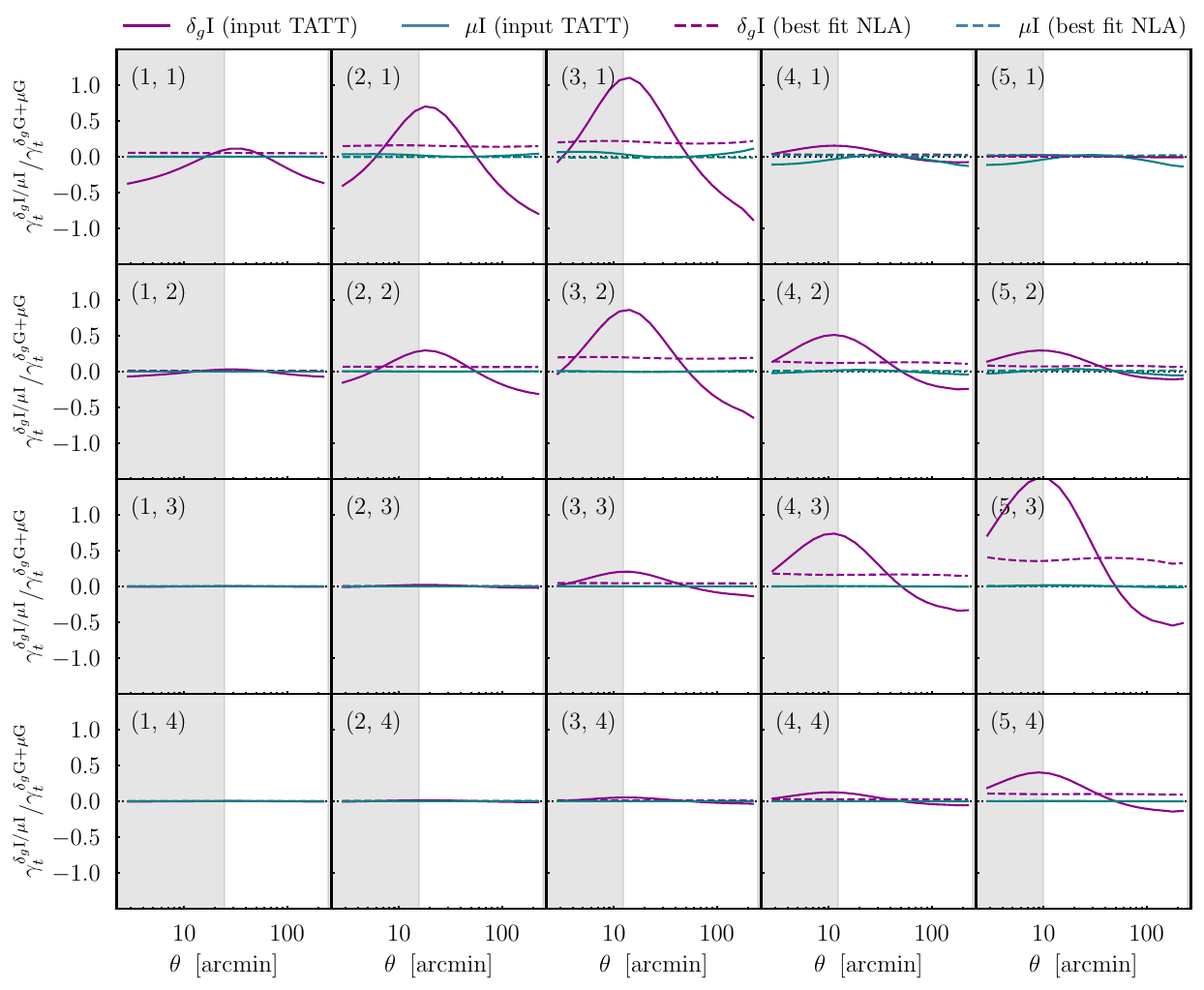}
    \caption{The fractional contributions of the $\delta_g$I and $\mu$I (galaxy-intrinsic and magnification-intrinsic) IA terms to our galaxy-galaxy lensing data. Both are shown relative to the cosmological signal (the sum of galaxy-shear and magnification-shear contributions $\delta_g$G and $\mu$G). As in Figure \ref{fig:xipm_GI_II}, shaded grey bands represent scale cuts. The numbers shown in each panel specify a particular bin pair $(l,s)$. The solid lines show the input IA signal for our mock $\gamma_t$ data, and dashed lines show the best fit prediction from an NLA $2\times2$pt analysis of that data. Note we are showing the \blockfont{redMaGiC} lens configuration here. See Figure \ref{fig:gammat_gI_mI_maglim} for the equivalent of this plot using the \blockfont{MagLim} lens sample. As with cosmic shear, the overall patterns of where in the data vector IAs are significant hold when plotting the other IA scenarios discussed in Section \ref{sec:results:applications}.\vspace{0.5cm} }
    \label{fig:gammat_gI_mI}
\end{figure*}

In the above subsections we have discussed the robustness of our central result. During the course of the discussion, we touched on some of the possible mechanisms that may contribute to this finding. Now we will draw these strands together to address the question more directly. 

There are a few possible explanations for the observation that a cosmological analysis based on cosmic shear is substantially more sensitive to IA mismodelling than an equivalent $2\times2$pt one. First, it could be that the IA contribution to $\gamma_t$, as a fraction of the total signal, is just generically smaller than in $\xi_\pm$. On the face of it, this is plausible, since the power spectra entering $\gamma_t$ and $\xi_\pm$ are filtered through different kernels in $z$ and $k$. Scale cuts and point mass marginalisation also act to remove sensitivity to smaller scales in $\gamma_t$, reducing the significance of higher order IA terms. We can rule this out fairly quickly, however using some basic observations. Comparing Figures \ref{fig:xipm_GI_II} and \ref{fig:gammat_gI_mI}, which show the fractional contribution of IAs to our fiducial shear and $\gamma_t$ data vectors, we can see that there is a substantial contamination in both data vectors; the maximal fractional contamination (relative to the lensing signal) is in fact larger in $\gamma_t$ than in cosmic shear. This can also be seen in Figure \ref{fig:chi2_initial_bias}, which shows significant levels of non-NLA contamination in our $2\times2$pt data vectors. Note also that, as seen in the literature, $2-$ and $3\times2$pt data typically constrain IA parameters more tightly than $1\times2$pt (this holds for TATT model constraints as well as NLA; see e.g. our Figure \ref{fig:contours_main}; Figure 8 of \citealt{samuroff18}; Figure 8 of \citealt{y3-kp}). This is true in general, unless the lens and source samples are specifically constructed to avoid IA contamination (see e.g. \citealt{miyatake23}). Given this, the generic insensitivity theory seems unlikely.

A related idea is that there is internal cancellation of the terms contributing to the GI power spectrum, leading to the TATT contributions being smaller in galaxy-galaxy lensing than shear. 
In principle, if the signal is entirely controlled by $P_{\rm GI}$, cancellation can occur in a way that is not possible when $P_{\rm II}$ (which is sensitive to the squares of IA amplitudes) is relevant\footnote{Note that cancellation can in principle occur between II and GI power spectra. This is, however, more difficult to achieve in practice since the two IA components tend to affect different parts of the data vector (i.e. it is relatively rare to have bin pairs that are equally affected by GI and II contributions on a given scale).}. Referring back to Eq. \eqref{eq:tatt_gi}, we can see that if $A_1$ and $A_2$ have opposite signs (as is the case in our fiducial example), the first two and final terms act against each other. Such cancellation can only ever be partial, and will affect some scales/redshifts more than others, since $P_\delta$, $P_{0|0E}$ and $P_{0|E2}$ have different shapes (e.g. Figure 1 \citealt{blazek19}). It can, however, potentially work to reduce the amplitude of the IA signal. This is slightly different from the previous explanation -- cancellation would imply the \emph{actual} TATT signal in the data is more often than not small, but does not imply a lack of constraining power (i.e. higher-order IAs can have an impact, in theory, but in practice they tend not to). There are some reasons to think such cancellation is a plausible factor. Considering the constraints in the literature, the degeneracies between $A_1$ and $A_2$ are such that it is much more common to find a combination of amplitudes with opposite signs than the same (see e.g. Figure 8 from \citealt*{y3-cosmicshear2} and Figure 7 from \citealt*{y3-sr}). Note also that in Figure \ref{fig:contours_main}, the posterior on $A_1$ from $2\times2$pt is closer to zero than the input value. That said, it does not appear that this is the whole picture. As mentioned above, there is relatively large data vector level contamination seen in Figure \ref{fig:gammat_gI_mI}, which is inconsistent with the idea that cancellation is \emph{the} main driver of the relative robustness of the $2\times2$pt data. There is also a significant impact coming from the non-NLA terms, as illustrated in Figure \ref{fig:chi2_initial_bias}. Considering Figure \ref{fig:gammat_gI_mI} we should note that although the NLA fit has a low amplitude, this is likely at least partly because of limited flexibility of the model. That is, because NLA is approximating an input TATT signal that oscillates between positive and negative on different scales, rather than because the TATT signal is necessarily small. We also note that although the bulk of the IA scenarios considered in Section \ref{sec:results:applications} do have opposite-sign $A_1$ and $A_2$, we see no systematic difference between same-sign and opposite-sign cases. The $2\times2$pt bias is typically smaller than the $1\times2$pt bias in both. 

One alternative hypothesis is that although IAs contribute to $\gamma_t$ measurements at a significant level, the scale and redshift dependence of that signal (after scale cuts) is simpler than in cosmic shear. If we refer back to the equations in Section \ref{sec:theory:ias}, we see that $\xi_\pm$ is sensitive to two IA components, GI and II (or three if we count the II B-mode spectrum separately), each of which can deviate significantly from NLA. On the other hand, $\gamma_t$ responds only to the GI term -- Eq. \eqref{eq:C_gI} and \eqref{eq:C_mI} depend on $P_{\rm GI}$ only. Ultimately we only need to solve for one IA power spectrum, to good approximation. It is possible that the unmodelled TATT signal in galaxy-galaxy lensing is just closer to NLA, and so more easily absorbed. Not only this, but scale cuts in $2\times2$pt analysis tend to more cleanly remove small physical scales than those in cosmic shear (see the discussion in Section \ref{sec:results:impact_of_bias_model}). According to this argument, the worst of the higher order TATT contribution is removed by the cuts on $\gamma_t$, and the surviving signal is relatively simple, allowing NLA to adjust in $2\times2$pt more easily than $1\times2$pt. In practice, however, there is at least some evidence against this, at least as a full explanation. We see in Figure \ref{fig:contours_main} that $A_1$ shifts significantly in response to the unmodelled TATT terms (implying the presence of at least some higher order signal to be absorbed after the scale cuts), but the details of the IA model used do not seem to be important. As we saw in Section \ref{sec:results:model_dependence}, reducing the IA model complexity worsens the $\chi^2$ but does not greatly increase the bias in cosmology. In Figure \ref{fig:ia_model} we found that even removing the IA model completely does not shift our fiducial case by more than a fraction of a $\sigma$. We should note that it is still possible to achieve this if \emph{both} the higher-order TATT terms are well approximated by NLA and some form of cancellation leaves the overall IA amplitude close to zero (i.e. by chance there is near-perfect cancellation). This seems somewhat unlikely, however, given our other observations. We can see that there are some non-trivial $\theta$ and redshift scalings in the TATT contributions in Figure \ref{fig:gammat_gI_mI}. In the bins where there is contamination, it is not clear that NLA is approximating the TATT signal any better here than it is in the cosmic shear case in Figure \ref{fig:xipm_GI_II}. This is reflected in the roughly comparable $\chi^2$ per degree-of-freedom for the two probes, as quoted in Section \ref{sec:results:sample2}. We should also note that we see a similar pattern across our 21 IA scenarios. We find the $\chi^2$ per degree of freedom from our $2\times2$pt chains is frequently at least as large as that from a cosmic shear analysis of the same IA scenario. This, again, suggests that higher-order TATT terms are not simply well approximated by NLA.

Our final explanation, which we believe is the most significant factor, is that internal degeneracy breaking means that IA error in $2\times2$pt simply does not translate into a bias in cosmological parameters in the way it does for shear. To see this, compare Figures \ref{fig:xipm_GI_II} and \ref{fig:gammat_gI_mI}. Whereas $\xi_\pm$ is affected by IAs throughout the whole data vector, in $\gamma_t$ the signal is relatively cleanly contained to certain redshift bin pairs. As illustrated in Figure \ref{fig:xipm_GI_II}, the TATT model can quite easily produce a significant II signal, especially in the auto-bin pairs, accompanied by a GI signal with quite different redshift and scale dependence. IAs appear in almost all parts of the $\xi_\pm$ data vector at a level between $\sim 10\%$ and $40\%$ of the cosmological signal. If we now consider Figure \ref{fig:gammat_gI_mI}, we see that the IA signal is similarly non-negligible in $\gamma_t$. In fact, in several bins it accounts for at least as large a fraction as, or even larger than, in $\xi_\pm$ (up to around $90\%$ on intermediate scales in certain bins -- this in consistent with the strong, if biased, constraint on $A_1$ shown in Figure \ref{fig:contours_main}). Crucially, however, the IA signal does not appear uniformly across the $\gamma_t$ data vector. Unlike in the case of cosmic shear, there are pairs of redshift bins where the lens-source overlap is very small, and hence in these bins there is almost no sensitivity to IAs (e.g. the lower left corner of Figure \ref{fig:gammat_gI_mI}). It is likely that this allows a degree of internal degeneracy breaking -- whereas a change in $S_8$ or $\omegam$~enters all $\gamma_t$ bins in roughly the same way, an unmodelled IA signal, clearly, can only appear in the bin pairs that have some sensitivity to IAs. This explanation is backed up by the observations in Figure \ref{fig:ia_model} (i.e. we are not reliant on any degree of flexibility in the IA model to absorb error). It is also supported by the fact that switching from \blockfont{redMaGiC} to \blockfont{MagLim} lenses (i.e. allowing the lens redshift distributions to shift/widen, and thus complicating the clean distinction between bins affected by IAs or not) is one of the few changes seen to increase the IA-induced bias in $2\times2$pt. Finally, when considering the ensemble of IA scenarios in Section \ref{sec:results:applications}, we see no significant correlation between the level of contamination and cosmological bias for $2\times2$pt (compare the black and purple points in Figure \ref{fig:chi2_initial_bias}; see also Figure 4 of \citealt{campos22}, which similarly shows a strong correlation for $1\times2$pt). The cases with the worst $\chi^2$ are not the ones with highest parameter bias, and the scenario that gives largest bias does not have a noticeably larger contamination than most of the other samples. This, again, points to a lack of degeneracy between the model error and cosmology as the driving factor, rather than e.g. cancellation or error being absorbed by the IA model. 

In summary, then, we have identified four potential factors that could all lead to $2\times2$pt analyses being less sensitive to IA modelling error than cosmic shear: (a) galaxy-galaxy lensing simply being less sensitive to IAs than cosmic shear; (b) cancellation of different TATT contributions, leading to a smaller higher-order IA signal; (c) a less complex TATT signal, that can be more easily approximated by a relatively simple model; (d) degeneracy breaking between IAs and $S_8$ in $\gamma_t$, enabled by the fact that IAs are confined to certain bin pairs. Considering the evidence for each explanation in turn, we conclude that (d) is likely the most significant effect. It is worth noting, however, that they are (mostly) not mutually exclusive. For example, is possible that some level of overall cancellation occurs in TATT scenarios with opposite sign $A_1$ and $A_2$. Likewise the differences in sensitivity to physical scales will likely have some impact on the magnitude of higher order terms and how a particular TATT IA power spectrum translates into bias. These effects exist and tend to act towards the same end, even if they are not the primary factor.

\vspace{0.5cm}
\subsection{Extension to future surveys}\label{sec:results:lsst}

\begin{figure*}
    \centering
    \includegraphics[width=\columnwidth]{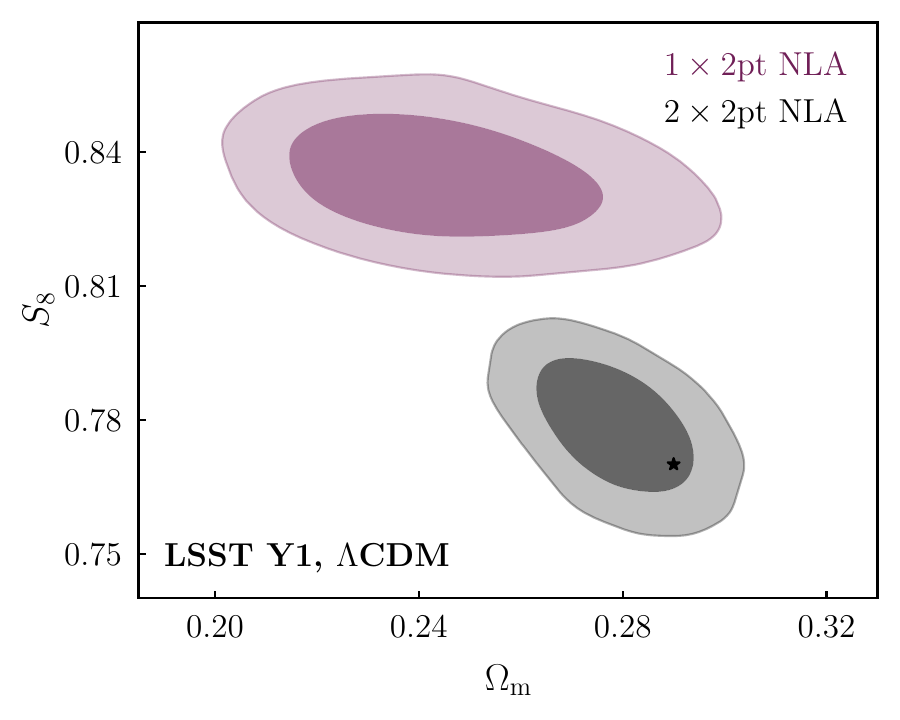}
    \includegraphics[width=\columnwidth]{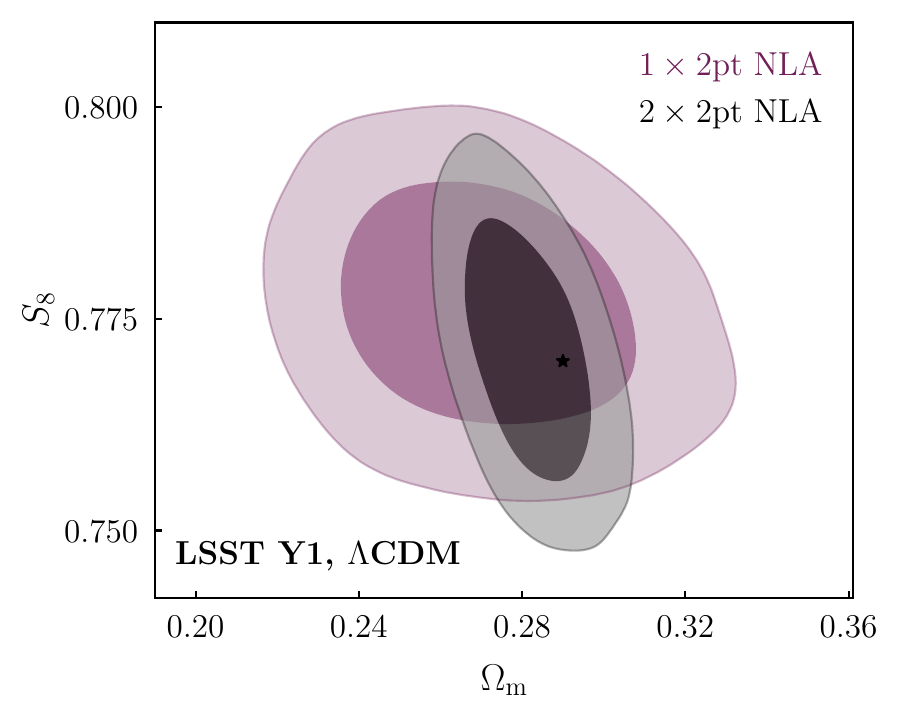}
    \includegraphics[width=\columnwidth]{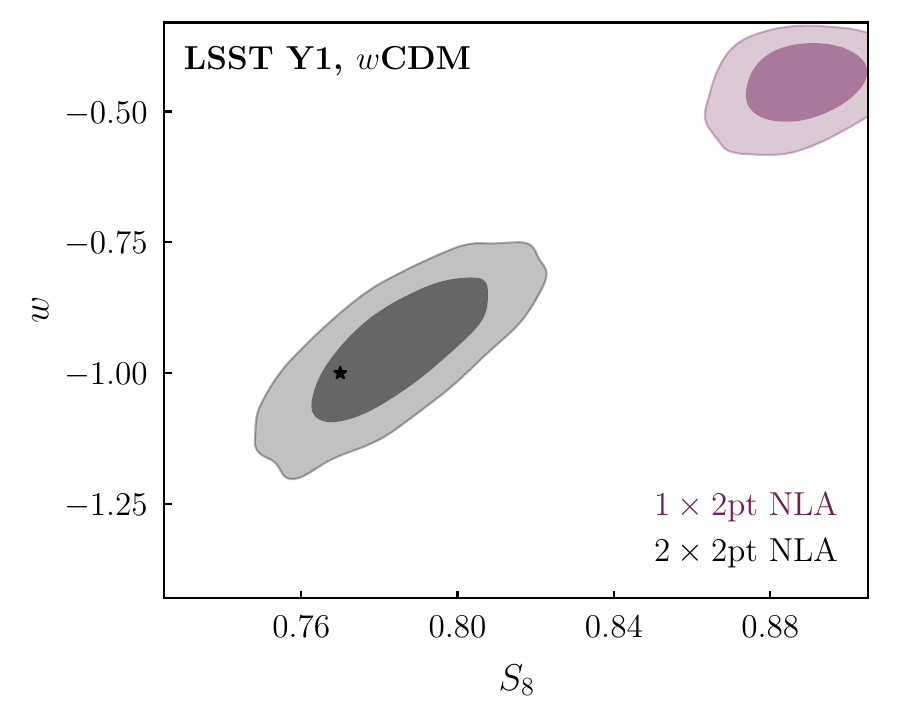}
    \includegraphics[width=\columnwidth]{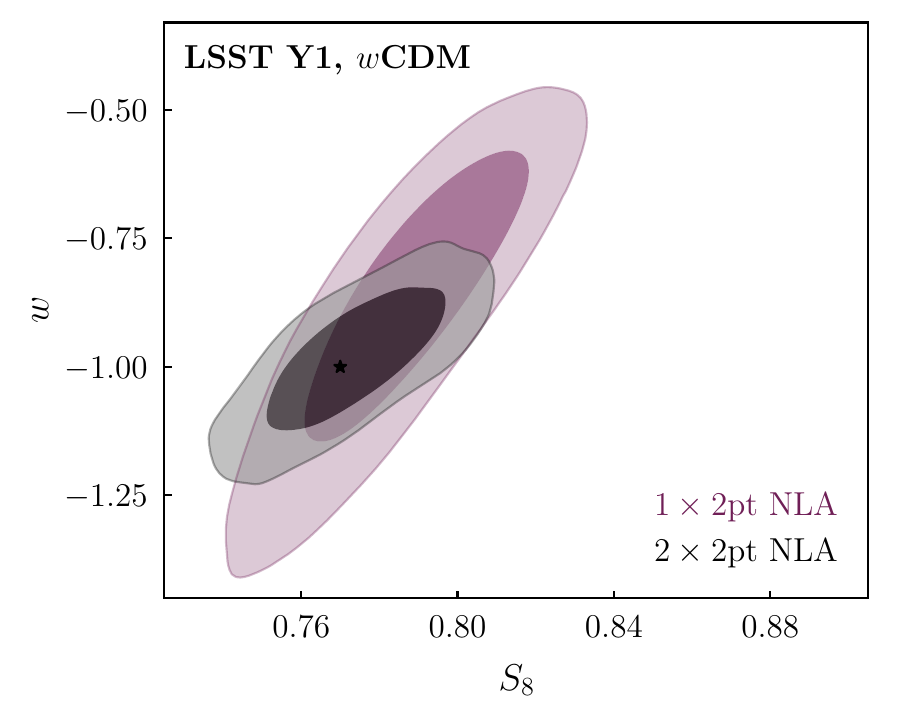}
    \caption{Predicted cosmological constraints from a mock Rubin Observatory LSST Year 1 like analysis. The left hand panels show the results from a data vector with a relatively extreme input IA scenario (the same as in Section \ref{sec:results:sample2}). The right shows a less extreme case with the input TATT parameters $(A_2,\eta_2)$ reduced. Both sets of input parameters are consistent at the $<2\sigma$ level with the DES Y3 results of \citet*{y3-cosmicshear2}; \citet{y3-cosmicshear1}. The upper and lower panels show \lcdm~and \wcdm~analyses respectively. In each case, purple contours represent cosmic shear only and black are galaxy-galaxy lensing + galaxy clustering. See Section \ref{sec:data:lsst} for details about the mock LSST data and analysis choices. As in previous figures, the black star marks the input cosmology.\vspace{0.5cm}}
    \label{fig:lssty1_posteriors}
\end{figure*}

The next question we consider in this section is how specific these findings are to our DES Y3 like setup -- i.e. can we extrapolate our conclusions to future data sets. To explore this, we use a set of Rubin LSST Y1 like simulations (see Section \ref{sec:data:lsst} and Appendix \ref{app:lsst_setup}). Compared with DES Y3, the constraining power is expected to increase significantly, reflecting an increase in the depth (i.e. number density) and area of both the source and lens samples (see e.g. \citealt{srd18}). We follow the DESC SRD by setting up our mock data with 5 source and 5 lens bins (shown in Appendix \ref{app:lsst_setup}), all of which have associated redshift uncertainties. We should note here that although these capture the rough gain in depth in LSST Y1, they are opimistic in the sense that they assume Gaussian uncertainties. In reality, redshift distributions estimated from real data tend to have non-trivial structure and tails, particularly at low redshift (Figure \ref{fig:nofz}; see also \citealt*{y3-sompz}; \citealt{giannini24}; \citealt{campos23}). These features increase bin overlap, and so tend to increase the sensitivity to IAs. Given this, we expect our LSST simulated tests to represent a lower bound on the impact of IAs on any future $2\times2$pt and cosmic shear analyses.

We choose two IA scenarios to focus on. The first is the same as in Section \ref{sec:results:sample2} above. This has not been ruled out by cosmic shear data to date, but is on the extreme end of what is consistent with previous results. In addition, we consider a less extreme case. For this, we keep $A_1$, $\eta_1$ and $b_{\rm TA}$ fixed, but scale down the tidal torque parameters to $A_2=-0.55$, $\eta_2=1.5$. This represents a reduction by roughly a factor of 5 in the amplitude of the quadratic IA term at the pivot redshift $z_0=0.62$; it also reduces the relative signal at high redshift, with an increase in $A_2(z)$ over the range $z=0.62 - 2.0$ of $\sim 2.5$ compared with $\sim 5.6$ in the more severe case. Note there is still a significant amount of uncertainty in the behaviour of IAs at higher redshifts, and these two values of $\eta_2$ are both entirely consistent with existing data. In the LSST setup, we run both \lcdm~and \wcdm~analyses. These are identical, except for the addition of a dark energy equation of state parameter $w$, varied with a flat prior over the range $w=[-2,-0.333]$.
Our findings are summarised in Figure \ref{fig:lssty1_posteriors}.

Unsurprisingly, we see a significant gain in the constraining power of each of the analyses compared with our DES Y3 like simulations (compare the contours in Figures \ref{fig:contours_main} and \ref{fig:lssty1_posteriors}). In \lcdm, like before, there is a significant bias in the NLA cosmic shear results (purple, labelled ``$1\times2$pt NLA") in the presence of IA modelling error. In the more extreme case (upper left panel), there is a shift of several $\sigma$ relative to the input in the high $S_8$ direction, very similar to the bias seen for our DES like setup in Figure \ref{fig:contours_main}. Note that the overall reduction in the size of the contours means there is less distortion due to prior edge effects -- we are no longer hitting the edge of the allowed parameter space in the $S_8-\omegam$ plane. In the less extreme scenario (upper right), the bias is small, at the level of $\sim0.5\sigma$ in the high $S_8$ low \omegam~direction. As before, we see that the galaxy clustering + galaxy-galaxy lensing results exhibit less bias than cosmic shear. Even in the case with larger unmodelled TATT parameters, the offset in the 2D $S_8-\omegam$ plane is below $1\sigma$ for $2\times2$pt. In the low TATT case, $S_8$ is almost unbiased, though there a small shift in $\omegam$. We note that for both IA scenarios shown here, rerunning these chains with TATT results in unbiased constraints (with projection effects at the level of $\sim 0.17 \sigma$ in the 2D $S_8-\omegam$ plane in \lcdm).

In the \wcdm~fits (the lower two panels of Figure \ref{fig:lssty1_posteriors}) we see a similar pattern. Interestingly, the trends here are even stronger than in the \lcdm. In the more extreme case (left), the $1\times2$pt posterior is shifted towards high $S_8$ and $w\neq-1$ by many $\sigma$. Here we \emph{are} beginning to hit the $w<-1/3$ prior bound (this is the reason for the difference in the size of the purple contour in the lower left and lower right panels). In contrast the $2\times2$pt constraint recovers the input to $<1\sigma$. Again in the weaker TATT scenario, there is a slight bias in the high $S_8$ - low $w$ direction, but it is sufficiently weak that the probes appear consistent.

Unlike in our DES Y3 like runs, we note that there is a very notable difference in the goodness-of-fit in these analyses. For the (we stress, noiseless) analyses of the extreme TATT scenario, we find $\chi^2_{\rm \Lambda CDM} = 334.8$ and $\chi^2_{w \mathrm{CDM}} = 312.2$ for cosmic shear. For $2\times2$pt, the equivalent best-fit values are much better at 10.6 and 6.8. We compute the number of effective degrees of freedom in the same way as in Section \ref{sec:results:sample2}, and find 277.6 and 277.5 for our \lcdm~and \wcdm~cosmic shear analyses respectively (286 data points). For $2\times2$pt, the equivalent values are 359.1 and 358.6 (378 data points). This gives typical $p-$values of $p(>\chi^2) \sim 10^{-27}$ for $1\times2$pt and $p(>\chi^2)\sim 0.3$ for $2\times2$pt. That is, given the precision of the Stage IV data, it seems clear that the goodness-of-fit would flag a problem in the $1\times2$pt chains. As we discussed in Section \ref{sec:results:sample2}, this was not the case for the same IA scenario at DES Y3 precision. In the less extreme case we find $\chi^2_{\rm \Lambda CDM} = 3.8$ and $\chi^2_{w \mathrm{CDM}} = 3.8$ for cosmic shear and $\chi^2_{\rm \Lambda CDM} = 7.9$ and $\chi^2_{w \mathrm{CDM}} = 6.9$ for $2\times2$pt.\vspace{0.1cm}

Interestingly, we note that our LSST $2\times2$pt setup experiences a weaker overall IA signal at the data vector level than the DES Y3 version of the same IA scenario (i.e. the equivalent to Figure \ref{fig:gammat_gI_mI} shows a smaller impact). This is reflected in the relatively good $\chi^2$ values listed above (despite the improved precision of the data). We put this down to a reduction in the fraction of the $\gamma_t$ data strongly affected by IAs, due to the highly idealised analytic redshift distributions in Appendix \ref{app:lsst_setup}. There are fewer bins with a strong IA signal compared with our real Y3 $n(z)$, and so the fit (and the resulting $\chi^2$) is dominated by lensing. 

Our overall conclusion from this exercise is that our qualitative findings from the previous sections hold, even with the enhanced statistical precision allowed by Rubin. That is, for a cosmic shear analysis in the presence of an unmodelled IA signal, model error tends to translate into potentially large bias in cosmological parameters. For a $2\times2$pt data vector affected by the same IA signal, in contrast, the posteriors are relatively well centred and the best fits are reasonably good.

\subsection{Exploiting internal tension as a sign of IA model error}\label{sec:results:tension}

\begin{figure}
    \centering
    \includegraphics[width=\columnwidth]{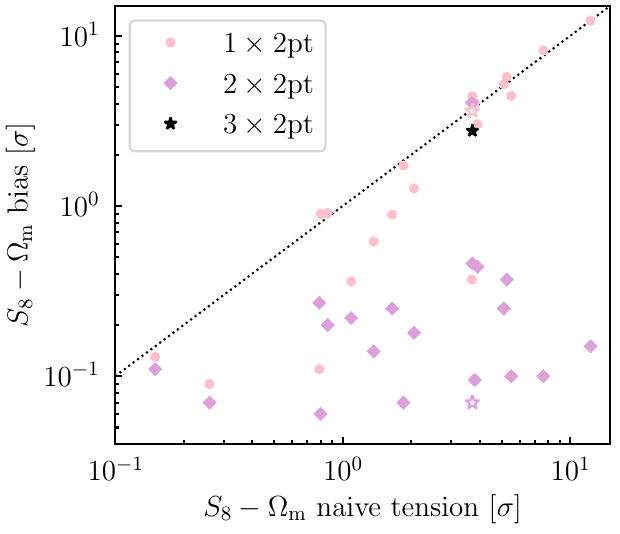}
    \caption{The relation between inter-probe ``tension" and cosmological bias in our 21 IA scenarios. Tension here is the distance between the peaks of the marginalised posteriors from cosmic shear and $2\times2$pt analyses of the same IA scenario. Bias is the offset of each from the input values. The diagonal line shows the 1-1 correspondence between the two. For illustrative purposes we also show the $3\times2$pt case for our fiducial IA scenario (the black star), which falls roughly halfway between the shear and $2\times2$pt results. As in Figure \ref{fig:bias_1x2pt_2x2pt}, the fiducial IA scenario is indicated as an unfilled star. \vspace{0.5cm}}
    \label{fig:bias_tension}
\end{figure}

Finally, we will briefly consider how IA driven differences between cosmic shear and galaxy-galaxy lensing might be used in practice. As we have seen in Section \ref{sec:results:sample2}, it is typically easier for model-data residuals from an insufficient IA model to translate into cosmological bias in a shear only analysis compared with $2\times2$pt. In Figure \ref{fig:bias_1x2pt_2x2pt} we see that IA scenarios that cause of the order of several $\sigma$ bias in cosmic shear can leave $2\times2$pt virtually unbiased. We should note that in practice, since the source sample used in shear and $\gamma_t$ measurements are generally the same, discrepancies of this magnitude are unlikely due to random fluctuations.

One consequence of this observation is that cosmological bias from IA model insufficiency can manifest itself as an apparent tension between probes. This could potentially provide a way of flagging an effect which can otherwise be difficult to detect using conventional goodness-of-fit and model comparison statistics (at least without calibration; see \citealt{campos22}). To illustrate this, Figure \ref{fig:bias_tension} shows the bias (i.e. the offset from the input truth in the $S_8-\omegam$ plane) plotted against the naive tension between shear and $2\times2$pt analyses (i.e. the distance between the peaks of the two posteriors). We show all 21 IA scenarios considered in Section \ref{sec:results:applications}. As we saw before, the $2\times2$pt bias falls consistently below the $1\times2$pt equivalent. Particularly in the high bias regime ($>1\sigma$) the ``tension" tracks the cosmic shear bias fairly well.

As referred to above, an offset in terms of number of $\sigma$ cannot be easily interpreted in terms of a probability when the data are correlated. There is typically some significant correlation between $\xi_\pm$ and $\gamma_t$, as measured from the same survey. In order to make a rigorous prediction one would need to use a metric designed to gauge these sorts of internal tensions \citet*{y3-internal_tensions}. That said, we might expect a more rigorous treatment of tension to shift the x-axis in Figure \ref{fig:bias_tension} to smaller values, but not necessarily to change the relation with bias. It is thus useful to consider this as a demonstration of the concept. 

\vspace{0.5cm}
\section{Conclusions}\label{sec:conclusions}

In this paper we have presented a series of simulated cosmological analyses, designed to shed light on the behaviour of joint weak lensing and clustering analyses in the presence of intrinsic alignment modelling error. Such joint analyses have become common in the field, and offer significant gains in terms of constraining power and self-calibration of uncertainties. In addition to those benefits, our results suggest that differences between probes in a shear + galaxy-galaxy lensing + galaxy clustering analysis may also provide a useful diagnostic. That is, we find that galaxy-galaxy lensing and cosmic shear have notably different sensitivities to IA modelling error. In the presence of a significant and insufficiently modelled IA signal, this can lead to apparent internal tension within the joint $3\times2$pt data vector. 
In more detail, our results are as follows:
\begin{itemize}
    \item At the precision of DES Y3, we find clustering + galaxy-galaxy lensing ($2\times2$pt) analyses to be relatively unbiased in the presence of a range of input IA scenarios ($<1\sigma$ in the $S_8-\omegam$ plane in almost all scenarios considered). This is true even when using a simple 1-parameter IA model. The same IA scenarios produce considerably larger biases in the equivalent cosmic shear analyses. Both the more-or-less unbiased $2\times2$pt fit and the considerably more biased $1\times2$pt one return acceptable $\chi^2$ values ($p-$values > 0.05).
    \item This robustness is not the result of insensitivity to IAs in galaxy-galaxy lensing: we find a significant contribution at the level of simulated data vectors. Rather, it points to a level of degeneracy breaking within the $2\times2$pt data vector, allowing even an overly-simple IA model to absorb the error rather than cosmological parameters.
    \item We find some dependence on the choice of lens sample. The differential sensitivity is most striking with a \blockfont{redMaGiC} like sample with well controlled photometric redshifts. Our conclusions also hold with a \blockfont{MagLim} like sample (greater number density, but wider priors on redshift nuisance parameters). We see slightly larger shifts in response to IA error, but still reliably under $1\sigma$ and much smaller than the equivalent shifts in cosmic shear. Similarly, including more small scale information in a $2\times2$pt analysis tends to increase the sensitivity to IA modelling error. Again, however, for all plausible setups tested, the bias was $<1\sigma$.  
    \item Considering the range of IA scenarios tested in \citet{campos22}, we find $>90\%$ are biased by less than $1\sigma$ in a DES Y3 like $2\times2$pt analysis (see Figure \ref{fig:bias_1x2pt_2x2pt}). The corresponding biases in cosmic shear span a range from $\ll 0.1\sigma$ to $>5\sigma$. This suggests that the differences are general, and down to a generic difference in sensitivity, rather than anything specific about our choice of fiducial input parameters.
    \item Extending our analysis to a mock LSST Y1 like setup, we find the same IA model error can, again, lead to differences of several $\sigma$ between cosmic shear and galaxy-galaxy lensing + clustering analyses. These differences are more pronounced in \wcdm~than \lcdm. We see a similar pattern as before, with much smaller biases in $2\times2$pt, and significant internal tension between $1\times2$pt and $2\times2$pt analyses based on the same survey.    
\end{itemize}

Considering a range of simulated tests and other evidence, we conclude that the most likely explanation for most of the differential sensitivity is internal degeneracy breaking, or self-calibration. Unlike shear, $\gamma_t$ measurements cleanly separate into bin pairs that are either strongly affected or more or less unaffected by IAs. This separation effectively breaks the degeneracy between $S_8 / \omegam$ and IA parameters. Other factors (e.g. signal cancellation and the greater control over the physical scales used in $\gamma_t$) likely contribute but are not thought to be the primary mechanism at work here.

In the wider context of the field, our results have a number of implications. Tests for internal tension have become a standard part of the unblinding process for many contemporary cosmology experiments \citep*{y3-internal_tensions}. Although conventionally thought of as a test for calibration errors and observational systematics, our results suggest they may carry information about modelling error as well. That is, the absence of internal tensions may be a sign that our choice of IA model is sufficient (at least at the level of $<\sim 1\sigma$).

Additionally, our results may help simplify the practical work of implementing data-driven model selection of the type proposed in \citet{campos22} (at least for IAs). If in general, cosmic shear is significantly \emph{more} sensitive to IA error than $2\times2$pt, it may be sufficient to perform model selection tests with the former. A calibrated goodness-of-fit that guarantees that our $1\times2$pt analysis is robust to modelling error at $<X\sigma$ (where we can choose our tolerance $X$), by extension will mean the same for $2\times2$pt.

Another implication is for the interpretation of IA constraints themselves. While the high S/N of a $\gamma_t$ measurement provides a potentially powerful constraint on IA parameters, that IA constraint is only as good as the model behind it. As we have seen, IA parameters can quite easily absorb modelling errors -- which, while useful for recovering cosmological parameters does tend to break the link with the physics of intrinsic alignments. Although this is true to an extent for cosmic shear as well, the tendency for IA errors to be absorbed by whichever IA parameters are available in $2\times2$pt analyses may make this more of an issue. As discussed in Appendix \ref{app:sr}, a similar thing can happen when shear ratios are included in a cosmic shear analysis, with model error being absorbed by the available parameters. 

In terms of the lens sample selection itself, our results offer a new angle. Although robustness to IA is unlikely to ever be the main driving factor, it is perhaps one consideration when constructing a lens catalogue -- that greater redshift precision translates into a $2\times2$pt analysis that is more robust to uncertainties in the IA model. Further along the pipeline, our results are also a motivation for including both very well-separated and overlapping source-lens bin pairs in the eventual cosmology analysis. As we have seen, the combination can provide enough information to disentangle IAs from cosmology relatively effectively.

We should also state that we are not presenting these results as a solution to the DES Y3 $X_{\rm lens}$ question \citep{y3-2x2pt_redmagic,y3-kp}. Tests have suggested $X_{\rm lens}$ was likely the result of selection-related systematics in the Y3 \blockfont{redMaGiC} lens sample, which can be avoided by changing the selection threshold. Although superficially similar to the sort of model-driven differences described here, the tension there was between bias values derived from $\gamma_t$ and $w(\theta)$. There are also features of $X_{\rm lens}$ (e.g. redshift dependence, the greater impact on \blockfont{redMaGiC} than \blockfont{MagLim}) that do not match well with the effects we describe in this paper.   

Finally, we highlight the importance of IA measurements of all sorts. There are several complementary ways to learn about IAs. Cosmological surveys contain more information about IAs than might be assumed. Most obviously, comparing the marginalised constraints from cosmological analyses provides a direct way to study the behaviour of IAs in real lensing data. Although this comparison can be done by reviewing the published literature, a deeper understanding can be obtained from analysing different lensing data sets in a homogenised model framework \citep{chang19,joudaki20,deskids23}. Further information can be gleaned about the sufficiency and relative favourability of different models from goodness-of-fit statistics (\citealt{campos22}; \citealt*{y3-cosmicshear2}). In addition to simultaneous constraints, direct measurements (e.g. \citealt{singh15,johnston19,fortuna22,samuroff22}) and tomographic IA-lensing separation (e.g. \citealt{blazek12}) 
are also key to fully understanding IAs in future lensing data sets. At present the conclusions we can draw from such measurements are limited by the data available. Future experiments such as the Dark Energy Spectroscopic Instrument (DESI; \citealt{desi16}), the Physics of the Accelerating Universe Survey (PAU; \citealt{serrano22}) and the Wide-Area VISTA Extragalactic Survey (WAVES; \citealt{driver19}), as well as subsets of data from LSST, Euclid and Roman, will offer us a way to test our models and begin to derive informative priors for IAs in lensing surveys. 

This paper provides one additional piece in the puzzle. We have shown that, since constraints from different probes respond differently to modelling error, a comparison of individual probe constraints can provide information that would not otherwise be obvious from the joint analysis of those same probes. In other words, there is value in analysing and presenting different probes separately. It is finally worth noting that, although we have attempted to test the robustness of our findings as much as possible, there are many factors at work. Dramatic changes in the source/lens samples or redshift binning, say, could affect our results. Likewise, an analysis with very different (or even no) scale cuts in either cosmic shear or galaxy-galaxy lensing (e.g. \citealt{arico23}) may respond differently. Taken at face value, the addition of very small scales in $\xi_\pm$ is likely to make one more sensitive to unmodelled higher order IA terms (and so increase the differences between shear and $2\times2$pt). Such analyses, however, tend to also include extra parameters to account for baryonic feedback, and it is difficult to predict how these will interact with higher order IA terms.

As data sets get deeper, we are also entering a regime where IAs are not well understood. As we have discussed in this paper, current data sets are primarily sensitive to IAs at low redshift. Upcoming surveys will reach much higher redshifts than any of the present generation, potentially significantly changing the sensitivity of the data to IA. Understanding the dynamics of IA constraints (i.e. the degree of potential self-calibration from different measurements, but also the possibility for greater sensitivity to error and for the failure of common parameterisations) is an important topic, and will be the focus of future work. 
\vspace{0.25cm}

\section*{Acknowledgements}

Thanks to Noah Weaverdyck and Johannes Lange for their help setting up the \blockfont{Nautilus} sampler. We would also like to thank our OJA reviewer for their constructive feedback. SS and JB are supported by NSF Award AST-2206563, the US DOE under grant DE-SC0024787, and the Roman Research and Support Participation program under NASA grant 80NSSC24K0088. AC acknowledges support from Department of Energy grant DE-SC0010118 and the NSF AI Institute: Physics of the Future, NSF PHY-2020295. AP acknowledges support from the UK Science and Technology Facilities Council (STFC) under grant numbers ST/V000594/1 and from the European Union’s Horizon Europe program under the Marie Skłodowska-Curie grant agreement 101068581. The contour plots shown in this paper were made using \blockfont{GetDist} \citep{lewis19}. This work was completed in part using the Discovery cluster, which is supported by Northeastern University’s Research Computing team.

\vspace{0.05cm}
\section*{Data Availability}

All simulated data vectors and chains used in this work are available on request.


\bibliographystyle{mnras}
\bibliography{refs}



\appendix

\section{Appendix A. Posteriors on redshift, galaxy bias and cosmology parameters}\label{app:other_params}
\begin{figure*}
    \centering
    \includegraphics[width=2\columnwidth]{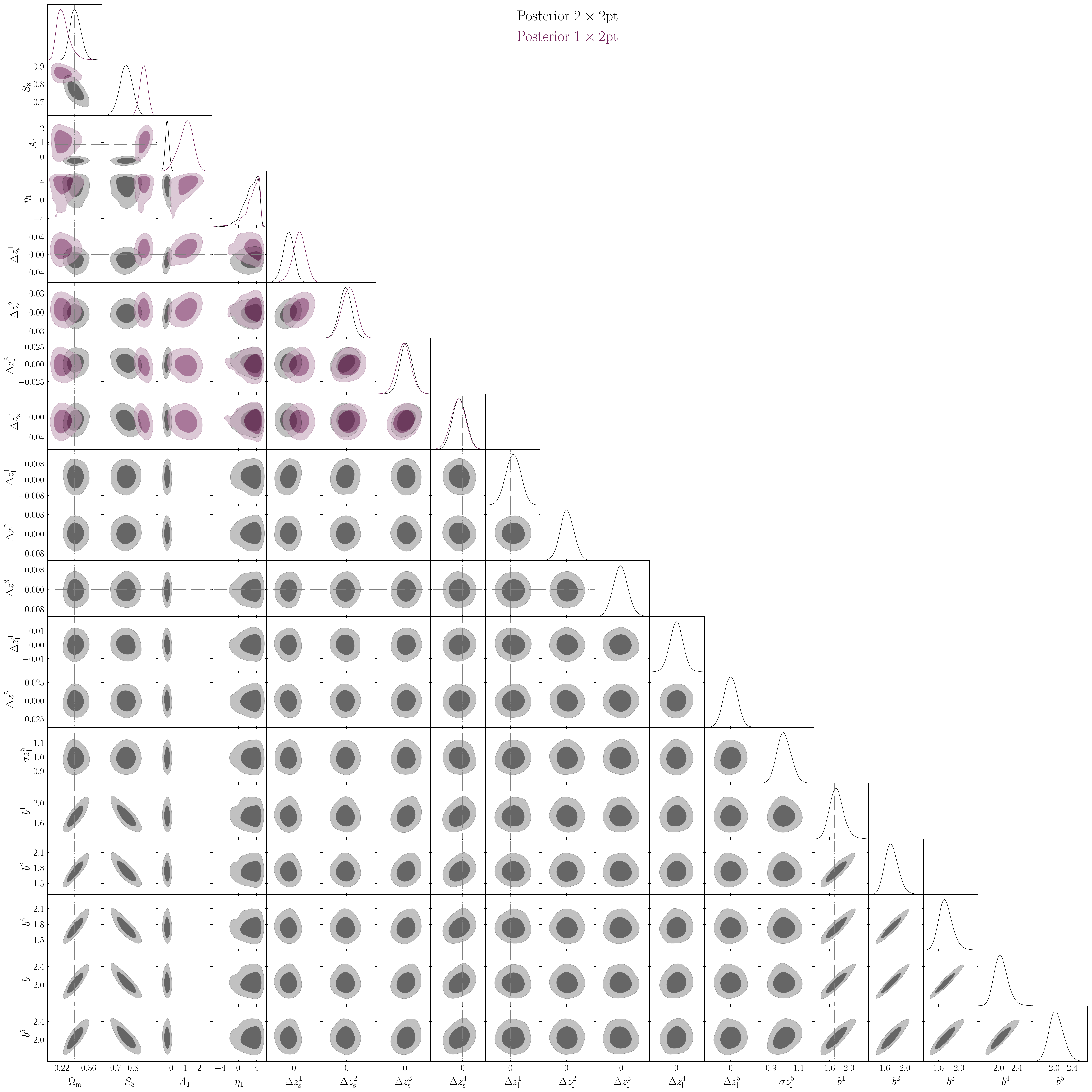}
    \caption{Posteriors on cosmology, redshift and galaxy bias parameters for the example IA scenario discussed in Section \ref{sec:results:sample2}. As in that section, we choose to focus on \blockfont{redMaGiC} lenses here, but a similar picture is seen with \blockfont{MagLim}. The purple and black contours show cosmic shear and galaxy-galaxy lensing + clustering analyses respectively. Although there are slight ($< 1\sigma$) shifts in the lowest source redshift bin (labelled $\Delta z^1_{\rm s}$ here), in general the source and lens redshift parameters and galaxy bias are stable and close to their input values (indicated by the dotted lines). The intrinsic alignment amplitude $A_1$, on the other hand, is seen to shift significantly in the $2\times2$pt case.\vspace{0.5cm}}
    \label{fig:giant_plot_redshifts}
\end{figure*}

\begin{figure*}
    \centering
    \includegraphics[width=2\columnwidth]{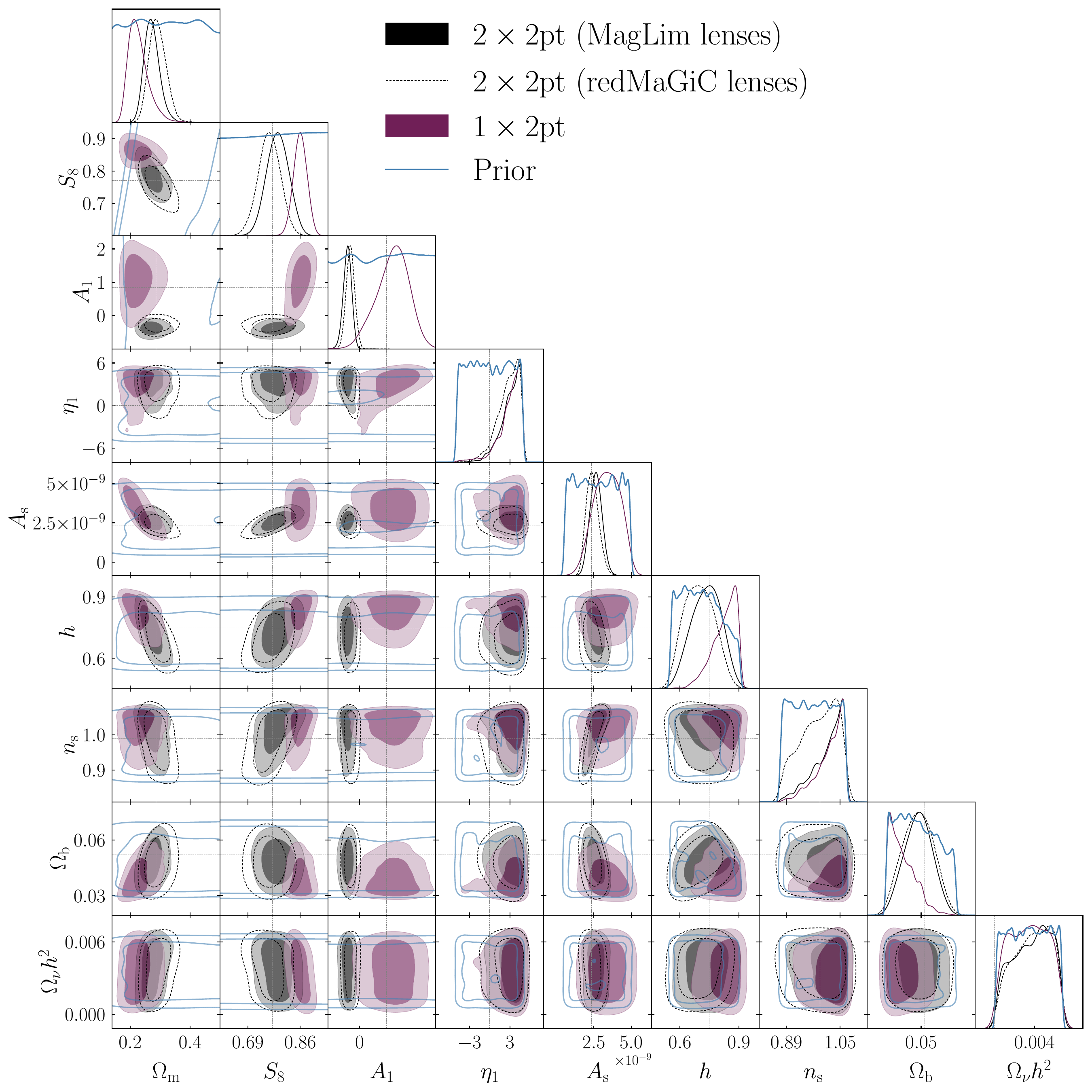}
    \caption{Posteriors on the full cosmology space from our mock cosmic shear only and galaxy-galaxy lensing plus clustering analyses. Overlain in blue is the prior (note this is a set of $\sim 10,000$ samples drawn from a $>15$ dimensional prior, and so there is some noise). The fine dotted lines in each panel show the input values used to generate the data vector. Although the cosmic shear only analysis (purple) shows some bias in this space, with $h$, $\ns$ and $\omegab$ pushing up against the prior edges, this is much less true for either $2\times2$pt posteriors (black shaded and dotted), which are largely unbiased.\vspace{0.5cm}}
    \label{fig:giant_plot_cosmology}
\end{figure*}

This appendix presents the posteriors in the parameter space of cosmology, IAs and redshift error parameters. 
Figure \ref{fig:giant_plot_redshifts} shows cosmic shear only (purple) and galaxy-galaxy lensing + galaxy clustering (black) analyses of our example data vector from Section \ref{sec:results:sample2}. The input parameters are indicated by the dashed lines. Although the IA parameters do shift, particularly in the $2\times2$pt case, the redshift parameters are noticeably stable. In neither analysis are either the lens or the source redshift nuisance parameters shifting to absorb the IA error present in the data. This is most likely at least in part due to the priors -- these are Gaussian and relatively informative, which restricts the ability of these parameters to vary too widely (see Table \ref{tab:params}). Note that we are using \blockfont{redMaGiC} here, but a similar plot using \blockfont{MagLim} lenses paints the same overall picture. Even with the extra width parameters associated with \blockfont{MagLim}, we do not see a significant amount of IA bias being absorbed into the redshift distributions. 

Figure \ref{fig:giant_plot_cosmology} shows the wider cosmology space. Overlain in blue is the prior. In the main body of the paper we focus on $S_8$ and \omegam, since these are the most constrained parameters. It can also, however, be interesting to look at the behaviour of the other, less constrained, parameters. Again, we do not see an obvious explanation for the relative lack of $S_8$ bias in the $2\times2$pt case. The neutrino mass parameter $\Omega_\nu h^2$ is almost completely unaffected, being strongly prior dominated. In general, $2\times2$pt (black filled and dashed) is in fact slightly more stable throughout the parameter space than cosmic shear (purple), which is shifted towards the prior edge in $\omegab, \ns$ and $h$. 

The overall picture here is fairly simple -- $2\times2$pt appears to be less biased in general than $1\times2$pt, throughout the whole cosmological parameter space, and not just the projected $S_8-\omegam$ plane. Interestingly the IA modelling error appears to be absorbed primarily by the IA parameters, without major leakage into other parts of parameter space.

\vspace{0.5cm}
\renewcommand{\thesection}{\Alph{section}}
\section{Appendix B. Comparing lens samples}\label{app:lens_samples}

In this appendix we consider in more detail the differences between our \blockfont{redMaGiC} and \blockfont{MagLim} lens samples, and the impact those differences have on our results.

Since the choice of lens sample has some effect on the level of cosmological bias in a NLA-based $2\times2$pt analysis (see Figure \ref{fig:contours_lenses}), it is worth trying to unwrap the differences. That is, going from \blockfont{redMaGiC} to \blockfont{MagLim} groups together a few different changes, and it is useful to consider them separately. 
The major differences are (a) the number of lens bins (5 for \blockfont{redMaGiC} and either 4 or 6 for \blockfont{MagLim}); (b) the shape of the estimated $n(z)$ themselves (c) the total effective number density $n_{\rm eff}$ of our \blockfont{MagLim} lens sample is greater than in \blockfont{redMaGiC} ($n_{\rm eff}=0.72$ galaxies/square arcmin and 0.17); (d) the uncertainties on the redshift estimates are larger for \blockfont{MagLim} (due to the inclusion of fainter galaxies), and so the width of the priors on redshift nuisance parameters is slightly larger.
We consider each of these in turn below.

Beginning with (a), the number of lens bins clearly has an impact on constraining power. This can be seen in Figure \ref{fig:contours_lenses} -- the pink contour is noticeably smaller than the purple. Including the upper two \blockfont{MagLim} bins clearly helps to constrain cosmology. It does not, however, shift the contour back to sit on top of the black (\blockfont{redMaGiC}) one. Although the test is not perfect (since different bins have different redshift priors), it does imply that the choice of binning alone is not the main factor in the \blockfont{redMaGiC}-\blockfont{MagLim} differences.

\begin{figure}
    \centering
    \includegraphics[width=\columnwidth]{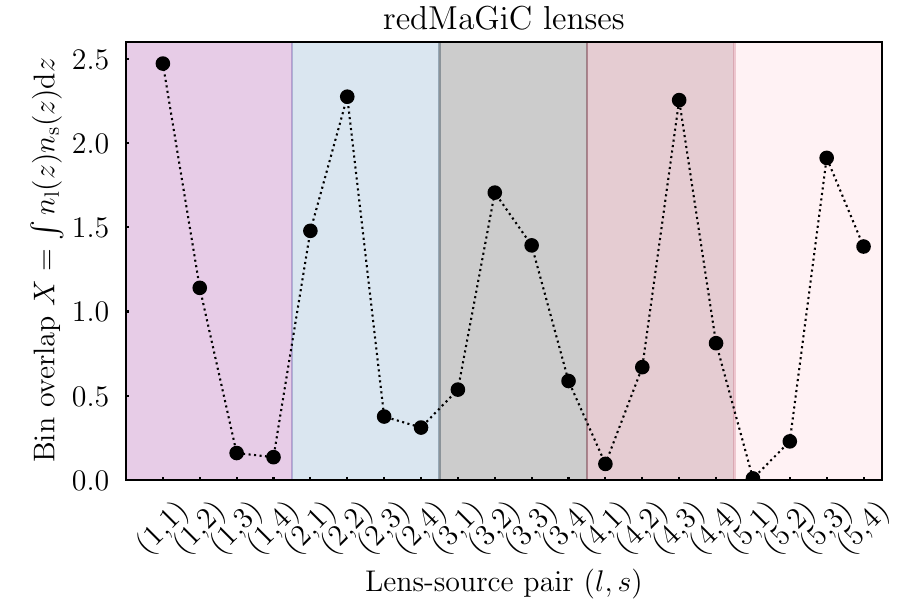}
    \includegraphics[width=\columnwidth]{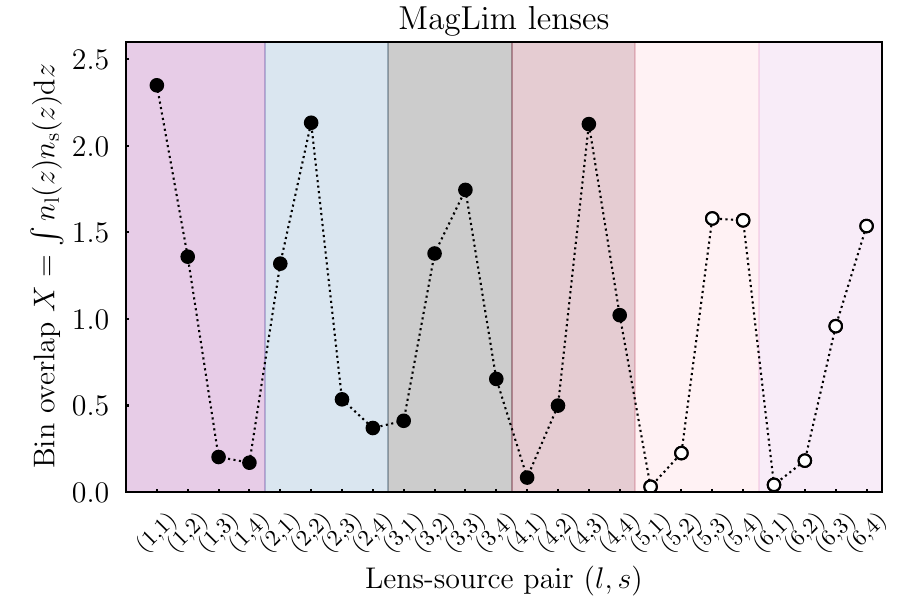}
    \caption{A quantification of the lens-source overlap for each possible bin pair. The quantity on the vertical axis is the total area enclosed by the product of the lens and source redshift distributions; this overlap term effectively modulates the size of the IA signal entering our $\gamma_t$ predictions (see the equations in Section \ref{sec:theory}). Each coloured band shows a particular lens bin. The upper and lower panels show the same calculation for \blockfont{redMaGiC} and \blockfont{MagLim} lens samples respectively. The overall pattern is seen to be roughly the same between the two lens samples. Pairs involving the upper two \blockfont{MagLim} lens bins are represented by open circles, as these are not included in the fiducial DES Y3 analysis.\vspace{0.5cm}}
    \label{fig:lens_source_overlap}
\end{figure}

Likewise we do not expect (b) to be the primary cause of differences in IA sensitivity. Although in principle differences in the shape and mean redshift of the lens distributions does alter the lens-source overlap, and thus the sensitivity to IAs, in practice the DES Y3 \blockfont{MagLim} and \blockfont{redMaGiC} $n(z)$ are not hugely different. This can be seen qualitatively by comparing the lower two panels of Figure \ref{fig:nofz}. A more precise comparison is given in Figure \ref{fig:lens_source_overlap}, where we show the integrated source-lens overlap, $X=\int n_{\rm source}(z) n_{\rm lens}(z) \mathrm{d}z $, which is the effective redshift kernel entering the galaxy-intrinsic ($g$I) contribution to $\gamma_t$. Again the differences are relatively small (compare the upper and lower panel; note that the y-axes are the same).
On the other hand, the combination of (c) and (d) \emph{is} thought to play a more significant part in the observed differences. These factors taken together are already known to have a significant impact on the $2\times2$pt cosmological constraining power (see \citealt{porredon21} Figure 10). That is, the increased density outweighs the slightly broader redshift priors, resulting in a tighter constraint in the $S_8-\omegam$ plane. This added density reduces the statistical uncertainty, and thus reduces the tolerance of small biases. Considering Figure \ref{fig:contours_lenses}, however, the difference between the black and purple contour is not simply a contraction about a shifted mean; a reduction in the uncertainty on $S_8$ by $10\%$ would not ultimately make much difference given the relatively accurate centring. 
The differences seen in Figure \ref{fig:contours_lenses} are thought to primarily arise from the broader $\Delta z$ and $\sigma z$ priors. In several bins we go from the $n(z)$ width being fixed entirely to being varied with a Gaussian prior (see the lower part of Table \ref{tab:params}). These width parameters are known to be degenerate with galaxy bias, and so with the overall amplitude of the $2\times2$pt data. By modifying the lens-source overlap, these parameters can effectively control how a given IA signal enters the data. This creates regions of parameter space where $S_8$ and IA error are degenerate, and so reduces the ability of different parts of the $\gamma_t$ data to self-calibrate such modelling error.

\begin{figure*}
    \centering
    \includegraphics[width=1.8\columnwidth]{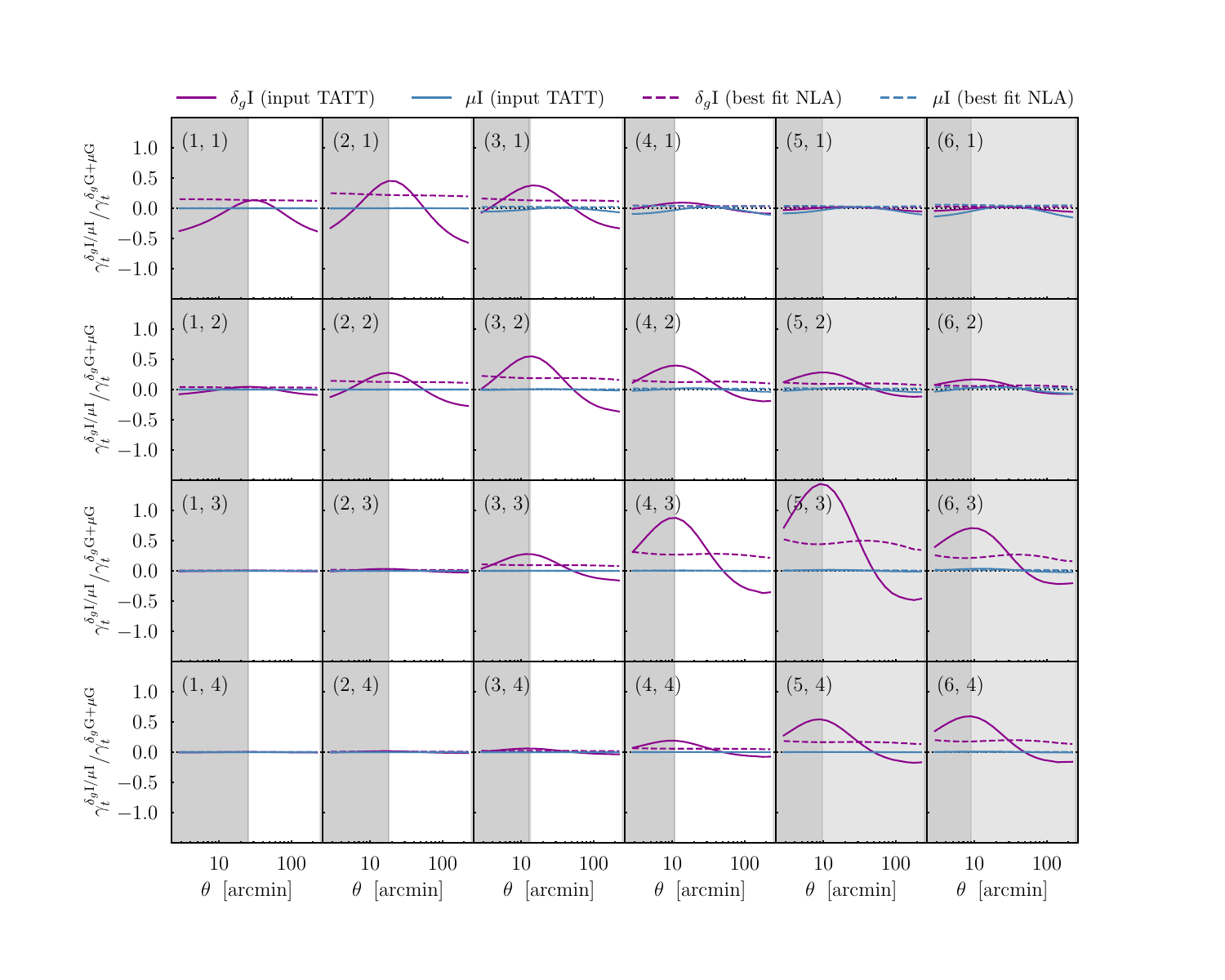}
    \caption{The fractional contributions of the $\delta_g$I and $\mu$I (galaxy-intrinsic and magnification-intrinsic) IA terms to our galaxy-galaxy lensing data. Both are shown relative to the cosmological signal (the sum of galaxy-shear and magnification-shear contributions $\delta_g$G and $\mu$G). As in Figure \ref{fig:xipm_GI_II}, shaded grey bands represent scale cuts. The numbers shown in each panel specify a particular bin pair $(l,s)$. The solid lines show the input IA signal for our mock $\gamma_t$ data, and dashed lines show the best fit prediction from an NLA $2\times2$pt analysis of that data. Note we are showing the \blockfont{MagLim} lens configuration here. See Figure \ref{fig:gammat_gI_mI} for the equivalent of this plot using the \blockfont{redMaGiC} lens sample.  \vspace{0.5cm}}
    \label{fig:gammat_gI_mI_maglim}
\end{figure*}

\vspace{0.5cm}
\renewcommand{\thesection}{\Alph{section}}
\section{Appendix C. Building an LSST Y1 like simulated analysis}\label{app:lsst_setup}

In this appendix we describe in more detail the details of our LSST Y1 like mock analyses. These are used in Section \ref{sec:results} to test the robustness of our results in an at least semi-realistic Stage IV like setup. The idea is not to predict every aspect of the future Rubin analysis correctly, but more to test if our findings are still applicable after a significant gain in constraining power. 

\begin{table}
	\centering	 
	\vspace{-0.2cm}
	\begin{tabular}{cccc}
		\hline
 Bin  & $n_{\rm eff}$ & Bias $b_1$ & $C$ \\
 \hline
  \hline
  1  & 3.60 & 1.42  & 0.43 \\
  2  & 3.60 & 1.66  & 0.30 \\
  3  & 3.60 & 1.70  & 1.75 \\
  4  & 3.60 & 1.62  & 1.94 \\
  5  & 3.60 & 1.78  & 1.56 \\
   \hline
\end{tabular}
 \caption{Lens sample parameters for our LSST Y1 like mock data. Shown (left to right) for each redshift bin are the effective number density, linear galaxy bias and magnification coefficient.}\label{tab:lsst_settings_lenses}
\end{table}

\begin{figure}
    \centering
    \includegraphics[width=0.9\columnwidth]{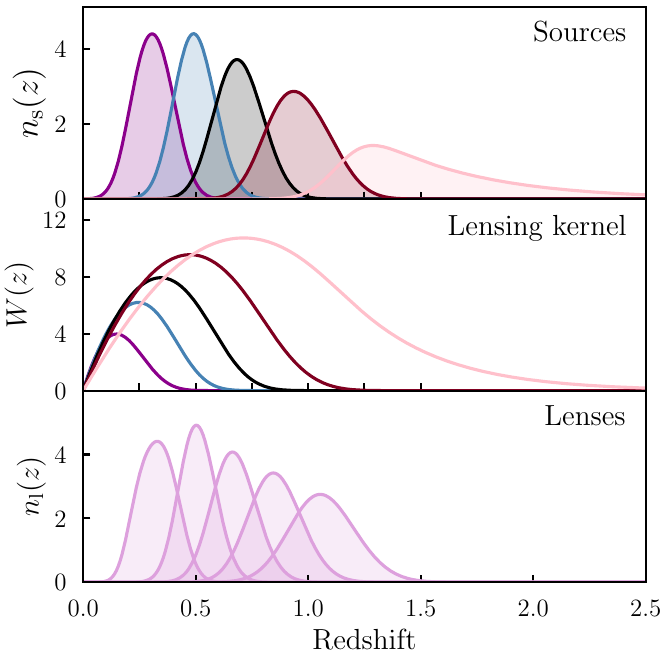}
    \caption{Redshift distributions for our mock LSST Y1 like samples. Note that these follow the prescription of \citet{srd18}. See Appendix \ref{app:lsst_setup} for discussion.\vspace{0.5cm}}
    \label{fig:lsst_nofz}
\end{figure}

\vspace{0.05cm}
For our LSST Y1 source sample, we follow \citet*{prat22} and define 5 equal number redshift bins. We assume the true underlying distribution has the form $p(z) \propto z^2 \times \mathrm{exp}[-(z/z_0)^\alpha]$, where $\alpha=0.78$ and $z_0=0.13$ (these numbers are taken from the DESC SRD - i.e. \citealt{srd18}'s Sec D2.1 and Appendix F2). A lower limit of $z>0.2$ is imposed, sharp bin edges are defined, and the resulting distributions are convolved with a Gaussian redshift error $\sigma_z=0.05$. This results in the source $n(z)$ shown in the upper panel of Figure \ref{fig:lsst_nofz} (which approximately match Figure 3 of \citealt*{prat22}). We follow \citealt{srd18} and \citet{fang20} in assuming a total number density of 11.2 galaxies per square arcminute, which divided across 5 bins, gives $2.24$ per bin. We also assume an ellipticity dispersion of $\sigma_e = 0.26$ in all bins.

For our lens sample, we assume five bins and again use an analytic form for the redshift true distribution ($\alpha=0.94$, $z_0=0.26$; see \citealt{srd18}). Following the predictions in the SRD, we truncate this distribution at $0.2<z<1.2$ and apply a Gaussian error of $\sigma_z = 0.03\times (1+z)$. The resulting lens $n(z)$ in five bins are shown in the lower panel of Figure \ref{fig:lsst_nofz}. We keep the Y3 \blockfont{MagLim} bias and magnification values as our inputs, as shown in Table \ref{tab:lsst_settings_lenses}. For the lens sample number density, we adopt the \citet{srd18} prediction for a Y1 photometric sample of 18 galaxies per square arcminute. Divided equally between our 5 bins, this gives us 3.60 per square arcmin. 

Note that in both cases, but particularly the source sample, these estimates will likely be quite different from the distributions eventually estimated from real LSST Y1 data. For one thing, the idealised selection function of \citet{srd18} will likely become more complicated, which could affect the depth as well as the shape of the distributions. Secondly, modern photometric redshift algorithms, designed to robustly characterise the ensemble $n(z)$ for a particular population (see e.g. \citealt*{y3-sompz}, \citealt{campos23}) typically result in non-Gaussian distributions with asymmetric tails to high/low redshift. Underestimating the tails is not necessarily an issue for predicting the shear signal alone, but it does matter for the IA contribution (sensitive to the shape of the lens and source $n(z)$). We should, then, consider our forecasts as a rough exercise to see how things change with greater depth and density rather than a robust prediction for LSST Y1.

Finally, we assume a joint lens-source mask with an area of 12,300 square degrees.
The joint covariance matrix of our LSST Y1 like shear, galaxy-galaxy lensing and clustering data is calculated using \blockfont{CosmoCov}. This computes both Gaussian shape noise and cosmic variance terms, as well as super sample and connected non-Gaussian contributions. Note this is the same methodology as was used for \citet{y3-kp}, albeit with updated parameters. 

For the analysis itself, we use the DES Y3 pipeline described in Section \ref{sec:theory}, with some differences. While source sample and cosmology priors are kept the same as in Table \ref{tab:params}, we update the lens sample priors to create a mock sample with \blockfont{redMaGiC}-comparable photometry. That is, we keep the width parameters fixed and adopt Gaussian priors on the $\Delta z$ parameters of widths 0.004, 0.003, 0.003, 0.005 and 0.01 respectively in our five lens bins.

For scale cuts, we apply the DES Y3 method of \citet{y3-methods}. That is, we generate mock LSST Y1 like data with and without baryonic contamination from the OWLS-AGN simulations \citep{vanDaalen11}. Scales are iteratively removed from the joint $\xi_\pm$ data vector until the total $\chi^2$ between the two data vectors $(\mathbf{D}^{\rm OWLS}-\mathbf{D}^{\rm fid}) \mathbf{C}^{-1} (\mathbf{D}^{\rm OWLS}-\mathbf{D}^{\rm fid})$ is below some threshold value. For the purposes of this work, we choose $\chi^2=2.5$ -- this is slightly less strict than the value eventually adopted by the DES collaboration \citep{y3-kp}, where the fiducial cuts were defined by $\chi^2=0.4$. There are two competing factors to account for here. The first, and perhaps more easy to predict, is that LSST Y1 will simply have more constraining power than previous surveys. This tightening of statistical uncertainties will allow less tolerance for a given degree of systematic error, and so will ultimately require more stringent scale cuts for any given model. The second is that there will inevitably be advances in the field of baryon modelling in the coming years. Even if it is not trusted on \emph{all} scales, it seems likely that Rubin will incorporate some sort of baryonic $P(k)$ model, and so access slightly smaller scales than they would if they were relying on a dark matter only model. It was this logic that led to our choice of 2.5. We should bear in mind that this is not designed as a robust prediction, but at least roughly accounts for these competing factors. For the galaxy-galaxy lensing and clustering parts of the data vector, we again follow \citet{y3-methods}. That is, we convert limits in comoving space at $6$ Mpc$/h$ (for $\gamma_t$) and $8$ Mpc$/h$ (for $w$) into angular scales using the mean redshift of each lens bin. We run two types of cosmological analyses in this LSST Y1 setup. The first is \lcdm, which uses the same priors and parameter choices as shown in Table \ref{tab:params}. The second is a \wcdm~analysis, which is the same, but with an extra equation of state parameter $w$, which is varied with a flat prior over the range $\mathrm{U}[-3,-0.33]$.

One consequence of the extra constraining power is that \blockfont{PolyChord} takes considerably longer to converge. We can understand this by considering how \blockfont{PolyChord}'s nested sampling works -- the prior distribution $P_\Pi$ sets a scale from which an initial set of live points are drawn. When the posterior is well-constrained compared with the prior volume, the algorithm can spend a significant time sampling from the tails of the distribution, meaning it is relatively slow to converge. This is necessary to accurately estimate the evidence. For this reason, for the LSST like chains we swap to \blockfont{Nautilus}. The basic idea of this new sampler is to use neural networks to help choose boundaries in parameter space most efficiently, and so minimise the total number of likelihood calls needed to estimate the posterior. The algorithm comes packaged with \blockfont{CosmoSIS} v3, and can be called in the same way as other samplers. We validate the convergence and accuracy of our LSST Y1 \blockfont{Nautilus} chains by re-running one of them using a more conventional MCMC approach -- \blockfont{emcee} \citep{foremanmackey12}. Specifically we re-run our less extreme IA scenario $1\times2$pt analysis (the purple contour in the upper right panel of Figure \ref{fig:lssty1_posteriors}). After allowing our \blockfont{emcee} chain to run for just over 1M samples, we cut the first quarter for burn-in. Plotting the surviving 750,000, we find good agreement with the shorter \blockfont{Nautilus} chain (the marginalised mean and standard deviation of $S_8$ are consistent to $<1\%$). Note that we do not use the Bayesian evidence values from \blockfont{Nautilus} in this work, so we do not compare or try to validate those here.

\vspace{0.5cm}
\renewcommand{\thesection}{\Alph{section}}
\section{Appendix D. The impact of shear ratios}\label{app:sr}

\begin{figure}
    \centering
    \includegraphics[width=\columnwidth]{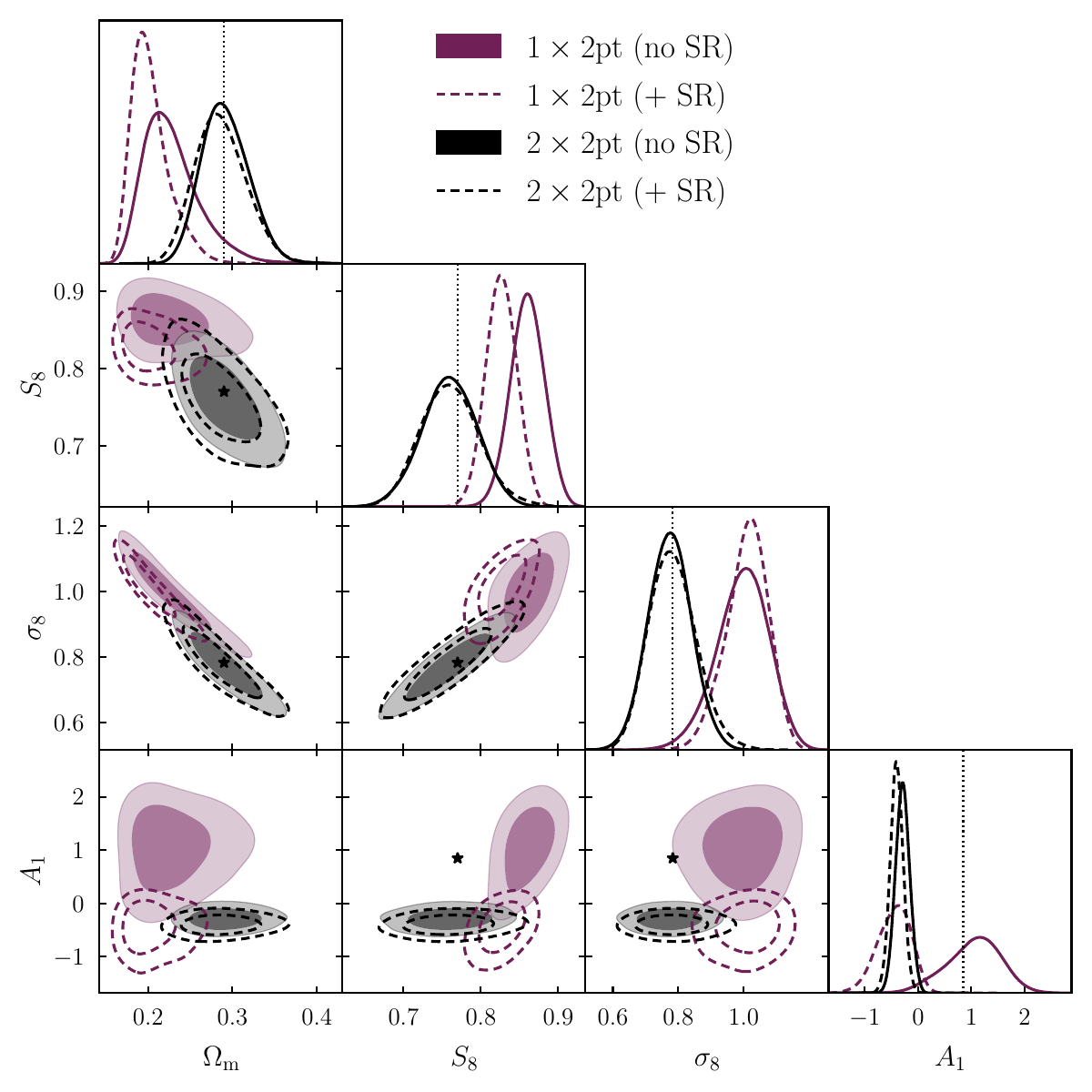}
    \caption{The impact of shear ratios. We show our fiducial $1\times2$pt and $2\times2$pt posteriors as shaded purple/black contours. The dashed contours are the same analyses, but including an additional likelihood from shear ratios on scales $2-6$ Mpc$/h$. Note that the simulated shear ratio data vector contains the same input IA scenario as the large scale $\xi_\pm$ and $\gamma_t$ data. As in previous plots, the input parameters are marked with a star. \vspace{0.25cm}}
    \label{fig:results:SR}
\end{figure}

In this appendix we briefly consider the impact of shear ratios (SR) on our results. SR are the ratios of galaxy-galaxy lensing measurements using the same lens bin. In theory the power spectra should (at least partially) cancel away, leaving ratios of angular diameter distances (see \citealt{jain04}, \citealt*{y3-sr}). For this reason they were proposed as a way to calibrate redshift error \citep{y1-ggl}. In practice, they can also contain a significant amount of information about intrinsic alignments (e.g. \citealt{blazek12}). The fiducial DES Y3 cosmic shear, $2\times2$pt and $3\times2$pt analyses (\citealt*{y3-cosmicshear2}; \citealt{y3-cosmicshear1}; \citealt{y3-2x2pt_maglim}; \citealt{y3-kp}) all included SR. We chose not to for the main sections of this paper because doing so mixes information coming from $\gamma_t$ and $\xi_\pm$, and so complicates our interpretation of differences between the probes. Since SR potentially carry information about IAs, however, it is interesting to briefly consider how our main results change as they are added. 

To avoid significant correlation with the large scale $\gamma_t$ and $\xi_\pm$ data, the DES Y3 setup includes SR measurements on scales of $2-6$ Mpc$/h$, where the uncertainties are shape noise dominated. Note that this is outside the one-halo regime, but roughly where higher-order TATT terms are expected to become significant (see \citealt{blazek19}). The covariance of the SR is propagated from the analytic covariance matrix used for the large scale inference. The SR data are averaged across angular scales, giving one data point per lens-source-source triplet. In total we have 9 data points from SR, constituting the first three lens bins, each paired with source bins (1,4), (2,4) and (3,4) \citep*{y3-sr}. The Y3 pipeline incorporates them as an additional likelihood, such that $\mathrm{ln} \mathcal{L} = \mathrm{ln} \mathcal{L}_{\rm 2pt} + \mathrm{ln} \mathcal{L}_{\rm SR}$ (see e.g. \citealt*{y3-cosmicshear2} Section IV-A).

To test how this extra information would interact with IA error, we set up a simulated test along the lines of those presented in Section \ref{sec:results}. Now, however, we also include a simulated SR data vector containing our fiducial TATT scenario (see Section \ref{sec:results:sample2}). We rerun our chains in this setup, modelling both the large scale two-point functions and the SR using NLA. We should note that this is a slightly contrived setup to help us understand how SR affect the dynamics of self calibration. The fiducial DES Y3 chains used TATT, which is valid on scales well below NLA.

Our results are shown in Figure \ref{fig:results:SR}. As we can see, the cosmic shear posteriors (the purple contours) are pulled to lower $S_8$. As we have discussed already, the direction is likely at least in part a function of the prior in the $S_8-\omegam$ plane, which restricts further movement in the low \omegam~direction. Interestingly the IA parameters are not only tightened considerably but drawn towards the region favoured by $2\times2$pt (the estimates of $A_1$, from cosmic shear + SR and $2\times2$pt with and without SR all agree very well, despite the information coming from very different scales). The end result is something not dissimilar to the joint $3\times2$pt results shown in Section \ref{sec:results:sample2} - the contours are tighter in the $S_8$ direction and shifted, but are still not correctly centred.

The $2\times2$pt contours on the other hand are stable, both in size and position. Interestingly, the impact is even smaller than in Figure \ref{fig:contours_bias_model}, where we saw a $\sim0.35\sigma$ shift when including scales in $\gamma_t$ down to $4$ Mpc$/h$. We can draw a few things from this. First, as seen in previous works (e.g. \citealt*{y3-sr}; \citealt{y3-cosmicshear1}) SR add most when IAs (or redshift parameters) are not well constrained. In the $2\times2$pt case, the large scale data provide a reasonable constraint, and so there is less impact in the $S_8-\omegam$~plane. Second, we see that $A_1$ changes very little when the SR likelihood is included. This is despite the fact that we are including scales where we expect the higher-order TATT contributions to worsen. One possibility here is that the IA constraints are simply dominated by the larger scales - this seems doubtful due to the relatively high precision of the SR data, and the fact that adding SR pulls the $1\times2$pt constraints on $A_1$ towards the same value. Another is that the TATT model on these scales can be consistently absorbed into the NLA model, implying a relatively weak scale dependence across the range $2-6$ Mpc$/h$. Finally, even with the addition of SR we see only small $<1\sigma$ bias in $2\times2$pt. We can take away from this that although we are including more contaminated data vector, on scales where the S/N is relatively high, the dynamic seen in a large scales only analysis is unchanged. That is, the data can still distinguish fairly well between shifts in $S_8$ and $\omegam$~and unmodelled IA contributions. The inclusion of SR in cosmic shear lessens the discrepancy slightly by shifting the cosmic shear posterior, but we still see IA modelling error manifested as a tension between probes. 

We should finally note that this setup mirrors the choices made in DES Y3. One could use SR purely as a redshift calibration tool by selecting bin pairs where the ratios are geometric (see \citealt*{y3-sr} Figure 2 and Section III). In this case, they will act only to tighten the posteriors on the (already largely unbiased) redshift parameters. Given this, we would expect even less impact compared with the solid contours in Figure \ref{fig:results:SR} in such an analysis. In short, our conclusions appear to be robust to the addition of small scale shear ratios.
\vspace{-1.5cm}

\vspace{2cm}
\end{document}